\documentclass[10pt, aps, prapplied, twocolumn, superscriptaddress]{revtex4-2}
\usepackage[english, french]{babel} 
\usepackage[T1]{fontenc}
\usepackage{hyperref, graphicx, xcolor, amsmath, amssymb}

\usepackage{bm, dcolumn}
\usepackage{braket, dsfont, textcomp, siunitx}
\hypersetup{colorlinks, linkcolor=teal, citecolor=blue, urlcolor=purple}

\usepackage{xcolor}
\usepackage{upgreek}
\usepackage{siunitx}

\newcommand{\LKB}{Laboratoire Kastler Brossel, Sorbonne Université, CNRS,
ENS-Université PSL, Collège de France, 4 place Jussieu, 75005 Paris, France}
\newcommand{\LPENS}{Laboratoire de Physique de l’École Normale Supérieure,
ENS, Université PSL, CNRS, Sorbonne Université, 75005 Paris, France}
\newcommand{\QUANTRO}{Quantronics group, Université Paris-Saclay, CEA,
CNRS, SPEC, 91191 Gif-sur-Yvette Cedex, France}

\begin{document}

\title{Hyperinductance based on stacked Josephson junctions}

\author{P.~Manset}
\affiliation{\LKB}
\author{J.~Palomo}
\affiliation{\LPENS}
\author{A.~Schmitt}
\affiliation{\LPENS}
\author{K.~Gerashchenko}
\affiliation{\LKB}
\author{R.~Rousseau}
\affiliation{\LKB}
\author{H.~Patange}
\affiliation{\LKB}
\author{P.~Abgrall}
\affiliation{\QUANTRO}
\author{E.~Flurin}
\affiliation{\QUANTRO}
\author{S.~Deléglise}
\thanks{These authors jointly supervised the work\\
Corresponding author: thibaut.jacqmin@lkb.upmc.fr}
\affiliation{\LKB}
\author{T.~Jacqmin}
\thanks{These authors jointly supervised the work\\
Corresponding author:
thibaut.jacqmin@lkb.upmc.fr}
\affiliation{\LKB}
\author{L.~Balembois}
\thanks{These authors jointly supervised the work\\
Corresponding author: thibaut.jacqmin@lkb.upmc.fr}
\affiliation{\LKB}

\date{\today}

\begin{abstract} Superinductances are superconducting circuit elements that combine a large inductance with a low parasitic capacitance to ground, resulting in a characteristic impedance exceeding the resistance quantum $R_Q = h/(2e)^2 \simeq 6.45~\mathrm{k}\Omega$. In recent years, these components have become key enablers for emerging quantum circuit architectures. However, achieving high characteristic impedance while maintaining scalability and fabrication robustness remains a major challenge. In this work, we present two fabrication techniques for realizing superinductances based on vertically stacked Josephson junctions. Using a multi-angle Manhattan (MAM) process and a zero-angle (ZA) evaporation technique — in which junction stacks are connected pairwise using airbridges — we fabricate one-dimensional chains of stacks that act as high-impedance superconducting transmission lines. Two-tone microwave spectroscopy reveals the expected $\sqrt{n}$ scaling of the impedance with the number of junctions per stack. The chain fabricated using the ZA process, with nine junctions per stack, achieves a characteristic impedance of $\sim 16~\mathrm{k}\Omega$, a total inductance of $5.9~\mathrm{\mu H}$, and a maximum frequency-dependent impedance of $50~\mathrm{k}\Omega$ at 1.4~GHz. Our results establish junction stacking as a scalable, robust, and flexible platform for next-generation quantum circuits requiring ultra-high impedance environments. \end{abstract}

\maketitle

\section{Introduction}

Superinductors and hyperinductors are an emerging new class of passive two-port superconducting elements whose performance is set by the following figures of merit:  their total inductance $L_\mathrm{tot}$ and their characteristic impedance $Z_c=\sqrt{L_\mathrm{tot}/C_\mathrm{stray}}$, where $C_\mathrm{stray}$ is the device's stray capacitance to ground. The super-inductance regime, defined by $Z_c>R_\mathrm{Q} = h/(2e)^2 \simeq 6.45$~k$\Omega$, combines a reactance well above the superconducting resistance quantum $R_\mathrm{Q}$, with perfect DC conduction and low microwave loss. This high-impedance environment freezes out charge fluctuations, while allowing for a continuous superconducting phase, paving the way towards novel quantum circuits architectures, including new superconducting qubits~\cite{manucharyan_fluxonium_2009,pop_coherent_2014,somoroff_millisecond_2023,koch_charging_2009}, proposals for protected qubits, like the $0-\pi$ qubit~\cite{kitaev_protected_2006,brooks_protected_2013,groszkowski_coherence_2018,gyenis_experimental_2021} and bifluxon~\cite{kalashnikov_bifluxon_2020,bell_spectroscopic_2016}, or alternative error-correction protocols~\cite{gottesman_encoding_2001,nathan_self-correcting_2024,sellem_dissipative_2025}. Beyond quantum computing application, a super-inductor can also enable the design of more compact resonators~\cite{ranni_high_2023}, that exhibit large zero-point voltage fluctuations. These resonators can be leveraged to enhance coupling to systems with weak dipole moments, such as quantum dots~\cite{bottcher_parametric_2022,stockklauser_strong_2017,landig_coherent_2018,scarlino_situ_2022}.

Several technological approaches currently coexist to realize super-inductors. The first one relies on the large kinetic inductance of thin, disordered superconducting films. Various materials can be used, such as NbN~\cite{niepce_high_2019}, TiN~\cite{shearrow_atomic_2018,peltonen_hybrid_2018}, or granular aluminum (GrAl)~\cite{grunhaupt_granular_2019,gupta_low_2024}. Historically, kinetic inductance has been considered the only viable mechanism to reach the super-inductance regime. However, this view has recently been challenged by the development of a geometric super-inductance technology based on concentric loops of aluminum nanowires~\cite{M_Peruzzo_2020,peruzzo2021geometric}. Finally, a third approach uses long chains of Josephson junctions, each operating well below its critical current to remain within the linear inductive regime. Additionally, if the Josephson junctions of such system exhibit a large Josephson‑to‑charging energy ratio ($E_J \gg E_C$), coherent quantum phase slips (CQPS)~\cite{randeria2024dephasing} are exponentially suppressed, even in the high‑impedance regime~\cite{masluk_microwave_2012,Manucharyan2011}.

This latter technology has recently been pushed to its extreme, enabling access to the hyperinductance regime~\cite{pechenezhskiy_superconducting_2020}, where the impedance of the inductive element exceeds $R_Q$ by at least an order of magnitude. This advancement has led to superconducting qubits in which the phase is a continuous variable spanning all real values. To reach extreme impedance, Pechenezhskiy et al. \cite{pechenezhskiy_superconducting_2020} have relied on suspended Josephson junction chains, elevated several micrometers above the substrate to replace the dielectric environment with vacuum. This reduces the parasitic capacitance $C_\mathrm{stray}$ by a factor $\epsilon_r / 2$ where $\epsilon_r$ is the relative permittivity of the substrate. While this method successfully enhances the characteristic impedance by a factor of $\sqrt{\epsilon_r / 2}$, it entails a technically demanding and mechanically fragile fabrication process with low scalability.

\begin{figure}[t]
    \includegraphics[width=0.46\textwidth]{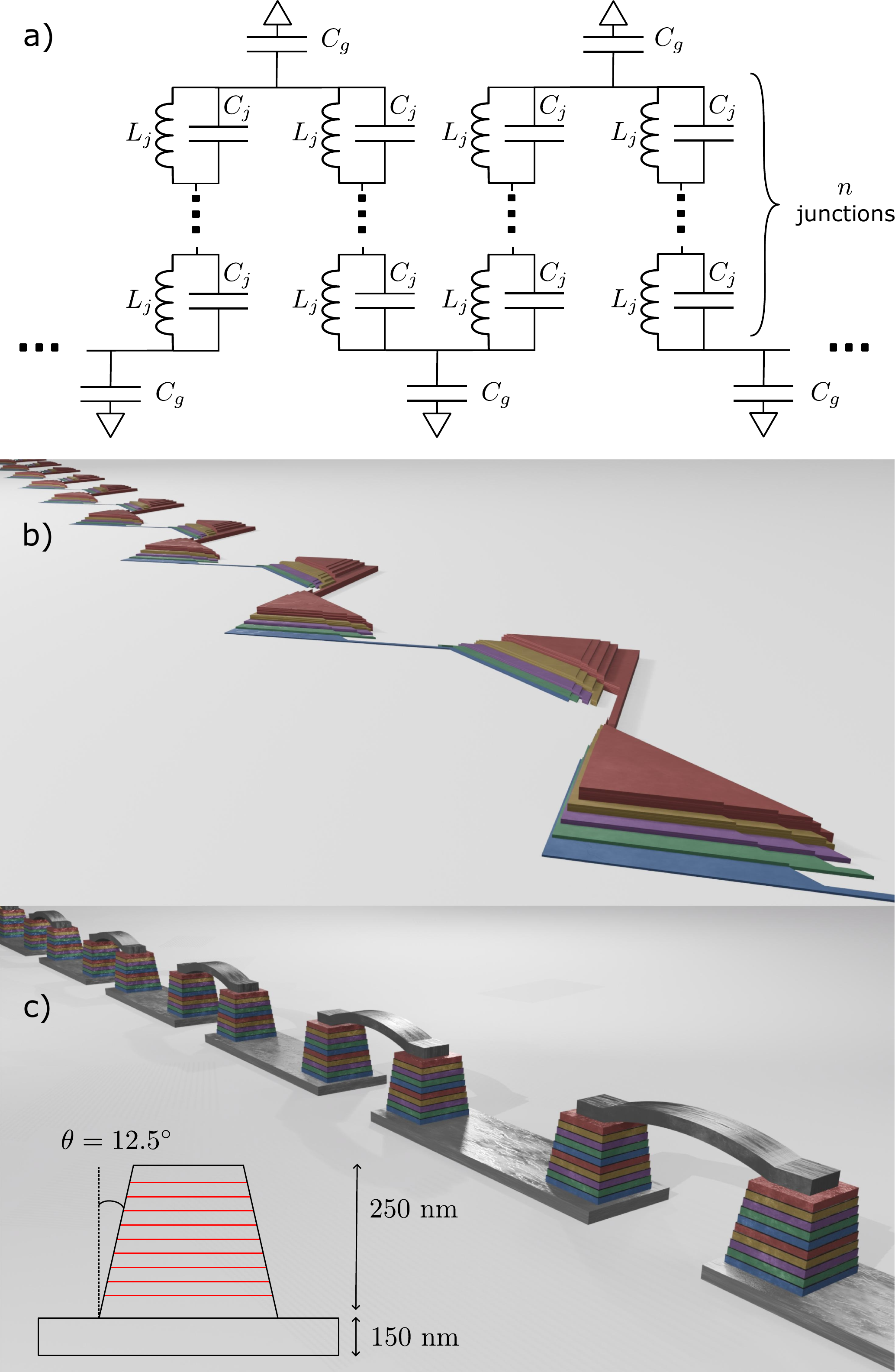}
    \caption{a) Lumped element model of a chain of $N$ stacks containing $n$ junctions each b) 3D modeling of the MAM chain, showing triangular shaped junction stacked on top of one another. In each stack, the different colors correspond to the 5 different layers in the structure, realizing 4 Josephson junctions. c)  3D modeling of the ZA chain showing the principle of the vertical stacking technique : 10 colorized layers are displayed and the oxide layers forming the 9 Josephson junctions are represented by thin dark grey strips.  The inset shows a cross-sectional view of a junction stack, illustrating its pyramidal shape resulting from mask clogging during the evaporation process. For visual convenience, the drawings are not to scale.}
    \label{fig:blender}
\end{figure}

In this work, we propose an alternative and scalable route to hyperinductance based on vertically stacked Josephson junctions. By arranging the junctions into a compact vertical, rather than horizontal chain (see Fig.~\ref{fig:blender}), the footprint of the device is significantly reduced, allowing for larger inductance density, while at the same time lowering the stray capacitance to ground, allowing for a larger impedance. More precisely, we implement various geometries of one‑dimensional high-impedance super‑inductors consisting of $N$ identical stacks of $n$ Josephson junctions each. Each junction is modeled identically as a linear inductance $L_J$ in parallel with a capacitance $C_J$, and each stack is coupled to ground via a stray capacitance $C_g$ (see Fig.~\ref{fig:blender}a), so that the chain forms a high-impedance transmission line with characteristic impedance $Z_c$.

\begin{equation}
\label{Eq:Zc}
Z_c = \sqrt{\frac{nL_J}{C_g}}.  
\end{equation}

From the transmission line spectrum, we can infer the plasma frequency $\omega_p=1/\sqrt{L_JC_J}$, identified as the high‑frequency cutoff. We also extract the other relevant frequency $\omega_g = 1/\sqrt{n C_g L_J}$ from the low‑frequency cutoff (first eigenmode frequency)
\begin{equation}
\label{Eq:omega1}
\omega_1 \simeq\frac{\pi}{N}\omega_g 
=\frac{ \pi Z_c}{L_{\rm tot}}
\quad\Bigl(L_{\rm tot}=nNL_J\Bigr),
\end{equation} 
which can then be used to obtain the characteristic impedance $Z_c$. 

To ensure the chain behaves as a purely linear inductance, it should be driven at frequencies $\omega<\omega_1$. Increasing the characteristic impedance $Z_c$ therefore extends this linear regime to higher frequencies. To increase the characteristic impedance $Z_c$ while maintaining a constant total inductance $L_{\rm tot} = nN L_J$, two design strategies can be employed: (i) increase $L_J$ by thickening the oxide barrier, while reducing the number of junctions $nN$; (ii) reduce the total stray capacitance $C_{stray} = N C_g$ by decreasing $N$, while increasing $n$ to keep $nN$ constant.

Since the first approach is ultimately constrained by CQPS—which, via the Aharonov–Casher effect perturbatively shift the qubit frequency and convert offset‑charge fluctuations on the array islands into dephasing~\cite{randeria2024dephasing}—the clearest path to an ultra‑high‑impedance, high‑inductance element is to employ a single Josephson‑junction stack containing a very large number of junctions.

This article is organized as follows: in Section~\ref{sec:cricuit_design}, we present the design of the super-inductance circuits designs. In Section~\ref{sec:dc_measurements} we estimate the total inductance of each chain thanks to room-temperature measurements of their resistances. Then in Section~\ref{sec:spectroscopy}, we perform spectroscopy on three different chains composed of stacked Josephson junctions. Section~\ref{sec:losses} provides an estimation of the chain losses and Section~\ref{sec:discussion} presents our conclusions.

\section{Circuit design}
\label{sec:cricuit_design}

\begin{figure*}[t]
    \includegraphics[width=1.0
    \textwidth]{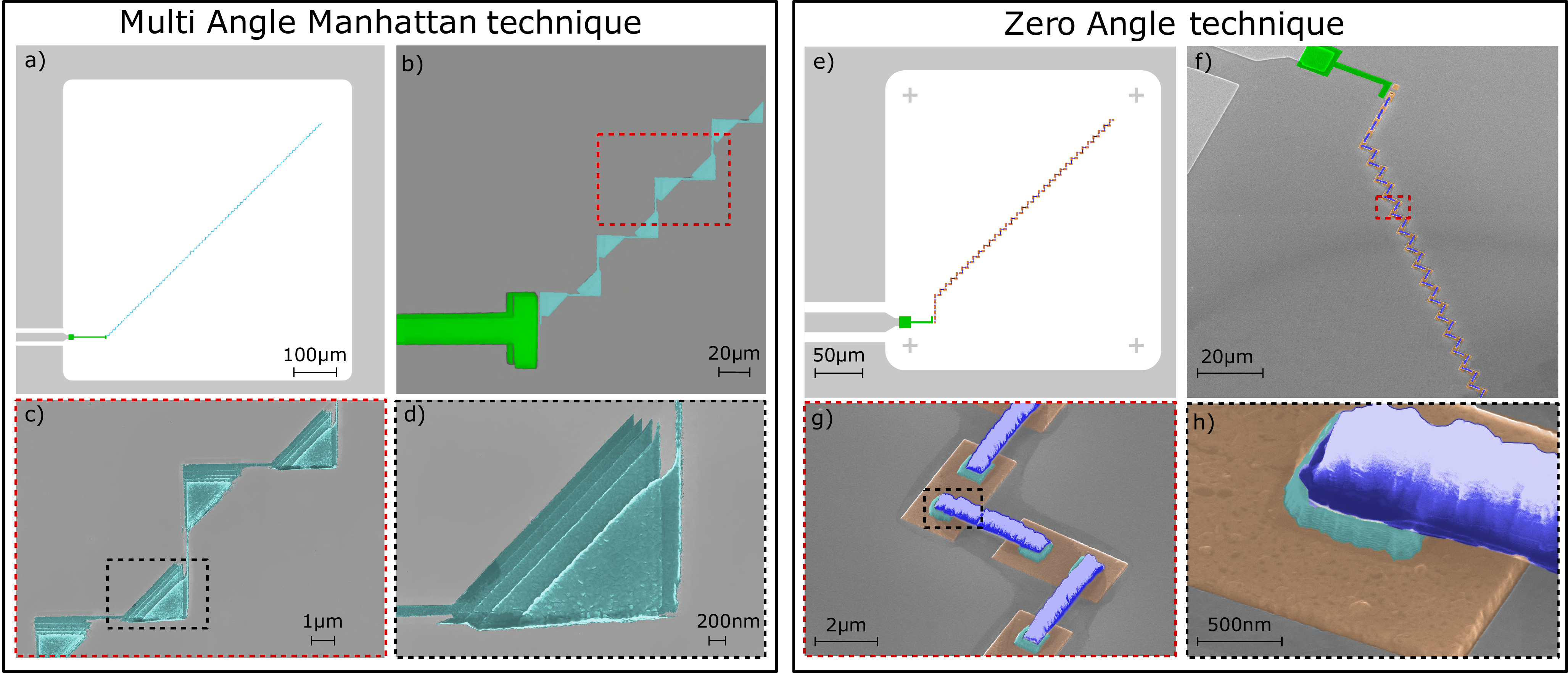}
    \caption{a) Schematic of the transmission line consisting of a chain of Josephson junction stacks (cyan), fabricated using the MAM technique.
    b) Optical microscope image of the chain. The aluminum coupling pad is colorized in green, and the chain of Josephson junction stacks is shown in cyan. The dashed red rectangle indicates the region magnified in c).
    c) SEM image showing four stacks, colorized in cyan. The dashed black rectangle indicates the region magnified in d).
    d) SEM image of a single stack containing nine Josephson junctions.
    e) Schematic of the transmission line consisting of a zigzag chain of Josephson junction stacks (red and blue), fabricated using the ZA technique and capacitively coupled to a microwave line. Grey planes indicate niobium layers, and the green structure represents the aluminum capacitive coupling electrode.
    f) Scanning electron microscope (SEM) image of the chain. The coupling pad is colorized in green, the air bridges are shown in blue, the bottom leads in orange. The dashed red rectangle indicates the region magnified in g).
    g) SEM image showing air bridges (blue) and bottom leads (orange) connecting stacks of junctions (cyan). The dashed black rectangle indicates the region magnified in h).
    h) SEM image of a single stack (cyan) containing nine Josephson junctions.
    }
    \label{fig:design}
\end{figure*}

We form a chain of Josephson junction stacks that is open-terminated on both ends (see Fig.~\ref{fig:design}a,e). This highly inductive line defines a $\lambda / 2$ resonator. The lumped element model of the stacked chain is similar to that of a simple single Josephson junction chain \cite{masluk_microwave_2012}, except that the junctions within a single stack now share a common capacitance to ground. 

We fabricated two types of junction chains. The first one is realized using a Multi-Angle Manhattan (MAM) evaporation process, in which zenithal and planetary evaporation angles are carefully controlled to define the entire stack without breaking vacuum. The stacks are composed of $n_{\textrm{M\!A\!M}}\!=\!4$ triangular junctions, whose areas range from \SI{1.8}{\micro\metre\squared} to \SI{2.1}{\micro\metre\squared} due to the various evaporation angles used in this technique. Moreover, these angles constrain the resonator shape to a zigzag geometry (see Fig.~\ref{fig:design}a,c) and won't allow for more than $\sim$ 6 junctions per stack due to clogging issues as detailed in Appendix~\ref{app:Fabrication}. The total number of stacks is $N_{\textrm{M\!A\!M}}\!=\!140$. A 3D rendering of the MAM structure is shown in Fig.~\ref{fig:blender}b, and a scanning electron micrograph is shown in Fig.~\ref{fig:design}c,d. To quantify the impedance gain of this method we also measured a reference chain with the exact same geometry but only one junction per stack, with areas of \SI{2}{\micro\metre\squared}.

The second type of junction chain is fabricated using a Zero Angle (ZA) technique, where $n_{\textrm{Z\!A}}\!=\!9$ junctions are stacked vertically using simple zero-angle evaporation, leading to  square-base pyramidal towers of height 250~nm, wherein the junctions' areas range from \SI{0.75}{\micro\metre\squared} to \SI{0.6}{\micro\metre\squared} due to the mask increasing clogging after each aluminum deposition cycle. The typical angle for this structure is $\theta = 12.5^\circ$ as shown in Fig.~\ref{fig:blender}c. The total number of stacks is $N_{\textrm{Z\!A}}\!=\!138$. The towers are galvanically connected two-by-two, alternatively from the bottom with aluminum stripes, and from the top with suspended aluminum bridges, usually referred to as \emph{air-bridges}. The bottom electrodes dimensions are \SI{4}{\micro\metre} $\times$ \SI{2}{\micro\metre}, and the aluminum bridge length is \SI{3.5}{\micro\metre}. A 3D rendering of this configuration is shown in Fig.~\ref{fig:blender}c, while a scanning electron micrograph showing a 9 junction stack on a bottom electrode, and contacted at the top by an aluminum suspended bridge is shown in Fig.~\ref{fig:design}g,h.
Note, that contrary to the MAM technique, the ZA method does not impose a zigzag shape to the line, and allows for arbitrary geometries.

The MAM and ZA devices are both capacitively coupled to a readout line on one end (see Fig.~\ref{fig:design}b,f bottom left L-shaped electrode).

\section{DC measurement of Josephson junctions stacks}
\label{sec:dc_measurements}

To verify that the inductance of a junction stack scales linearly with the number of constituent junctions, we perform room-temperature resistance measurements on dedicated junction chains. In a single Josephson junction, the link between resistance and inductance is given by the Ambegaokar–Baratoff relation \cite{ambegaokar1963tunneling}
\begin{equation}
    R = \frac{\pi \Delta}{2e I_c},
\end{equation}
where \( R \) denotes the room-temperature resistance, \( \Delta \) is the superconducting energy gap (of the aluminum in our case), and \( I_c \) is the junction critical current, which is inversely proportional to the junction inductance. As a result, the total room-temperature resistance of a junction chain is directly proportional to its total inductance. In thin films, the value of the superconducting energy gap tends to be higher than that of the bulk material \cite{ferguson2007energy}. We independently calibrate this discrepancy and account for it by applying an empirical multiplicative factor of \( 1.45 \) to the standard Ambegaokar–Baratoff relation.

\begin{figure}[t]
    \includegraphics[width=0.46\textwidth]{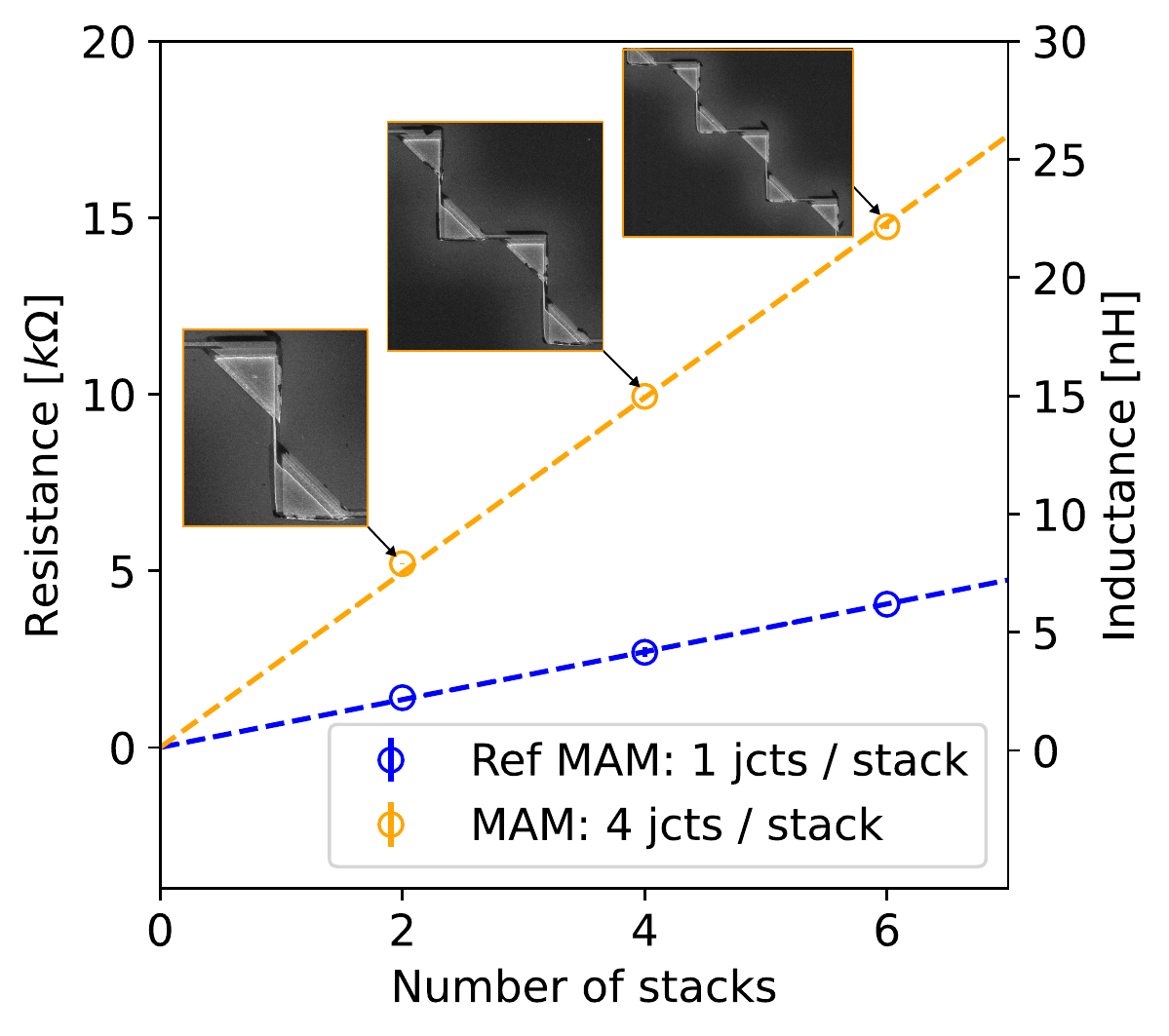}
    \caption{DC measurements of two chains fabricated using the MAM technique. Solid markers represent the experimental data, while dashed lines indicate linear fits constrained through the origin. Inset are SEM images of the stacks with 4 junctions per stacks. The y-axis on the right shows the inductance values inferred from the Ambegaokar–Baratoff relation.}
    \label{fig:dc_measurements}
\end{figure}

Each sample contains 18 replicas of test chains with 2, 4, and 6 stacks. We perform measurements on MAM chains, with one and four junctions per stack respectively. The average resistance measured for the two samples is shown in Fig.~\ref{fig:dc_measurements}. The standard deviation remains below \( 0.1~\text{k}\Omega \) for all measurements. The linear fit of the resistance as a function of the number of stacks yields a slope of 0.67 k$\Omega$/stack (respectively $2.48$ k$\Omega$/stack) for the samples with one (respectively four) junction(s) per stack. This confirms the expected factor of four between the two sample. 

Finally, by dividing the total inductance of the chain by the number of junctions it contains, we can extract the average inductance of a single junction,  $L_j = 0.96 \pm0.05$~nH.

\section{Chain impedance measurement via spectroscopy}
\label{sec:spectroscopy}

\begin{figure*}[t]
    \includegraphics[width=1.0
    \textwidth]{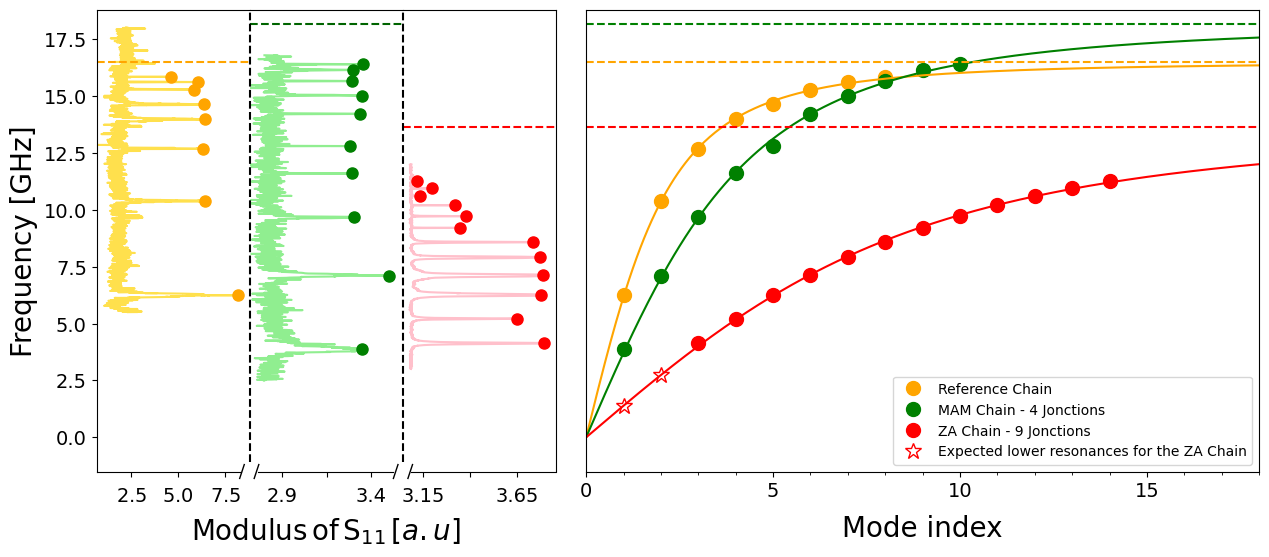}
    \caption{Two-tone measurements of the different structures. The three figures on the left show the magnitude of the reflection coefficient of the probe signal as a function of the pump signal frequency. From left to right, we show measurements made on the Reference, the MAM and ZA chains, respectively. Peaks, indicating resonant chain modes, are marked by dots and are fitted on the right panel with Eq.~\ref{Eq:eigenmodes}. Each horizontal dotted line highlights the plasma frequency in each chain according to the fit.}
    \label{fig:plasma_freq}
\end{figure*}
The resonator modes frequencies are determined by computing the total transmission matrix of the chain and applying open boundary conditions (see Appendix~\ref{app:theory}). We find
\begin{equation}
    \omega_m = \omega_p \sqrt{\frac{2\left(1 - \cos\frac{m\pi}{N}\right)}{\frac{\omega_p^2}{\omega_g^2} + 2\left(1 - \cos\frac{m\pi}{N}\right)}},
\label{eq:dispersion_relation}
\end{equation}
where \( m \) denotes the mode index. In the asymptotic limit \( \pi/N \ll \omega_p/\omega_g \ll 1 \), the spectrum separates into two regimes. The low-frequency modes exhibit a nearly linear dispersion, \( \omega_m \sim m \omega_1 \), where \( \omega_1 = \pi \omega_g / N \). In this regime, the capacitive energy is primarily stored in \( C_g \). Conversely, the high-frequency modes are predominantly localized in \( C_J\), and their frequencies asymptotically approach \( \omega_p \).
The slope of the linear dispersion can be expressed in terms of the chain's characteristic impedance and total inductance
\begin{equation}
    \omega_1 = \pi \frac{Z_C}{L_{\text{tot}}}.
    \label{eq:omega_1}
\end{equation}
If a sufficient number of modes can be resolved experimentally, both asymptotic behaviors can be extracted, providing independent estimates of \( \omega_g \) and \( \omega_p \). Since those frequencies are functions of three unknown parameters—\( L_J \), \( C_J \), and \( C_g \)—an additional input is required to fully characterize the system.
To address this, we rely on two complementary approaches: (i) estimating \(L_J\) from the room-temperature resistance measurements presented in section \ref{sec:dc_measurements}; and %on dedicated test junctions, which, %via the Ambegaokar-Baratoff relation~\cite{ambegaokar1963tunneling}, yield an estimate of \( L_J \); and 
(ii) electromagnetic simulations of the fabricated geometry, which provide an estimate of \( C_g \).

Since the mode frequencies \( \omega_m/2\pi \) span a range that extends both below and above the 4–12~GHz bandwidth of our amplification and measurement chain, not all modes are accessible via direct vector network analyzer (VNA) measurements. To overcome this limitation, we use two-tone spectroscopy \cite{masluk_microwave_2012}: a pump tone is swept across a broad frequency range while monitoring the frequency shift of a fixed probe mode within the accessible bandwidth. The pump amplitude is set high enough for the junction nonlinearity to induce a cross-Kerr shift on the probe mode, making it sensitive to the excitation of otherwise inaccessible modes.

Two-tone spectroscopy measurements for the Reference, MAM, and ZA chains are shown in Fig.~\ref{fig:plasma_freq}a. The resonant modes appear as sharp peaks in the amplitude response \( |S_{11}| \) of the probe tone and are identified using a peak detection algorithm (solid points). Due to the characteristics of the directional coupler used to route the pump tone to the device, modes below 3~GHz could not be resolved. While this low-frequency technical cutoff lies below the fundamental mode \( \omega_1 \) for the Reference and MAM chains, it falls within the spectrum of the ZA chain, requiring us to infer the index of the first visible peak for that device. The extracted mode frequencies as a function of mode index are presented in Fig.~\ref{fig:plasma_freq}b (solid points), along with fits to the model described by Eq.~\eqref{eq:dispersion_relation} (solid lines). 
The empty star symbols represent the frequencies of the first two missing modes, as inferred from the fitted parameters.

We extract plasma frequencies of \( \omega_p/2\pi = 16.50 \,\mathrm{GHz} \), \( 18.18 \,\mathrm{GHz} \), and \( 13.66 \,\mathrm{GHz} \) for the Reference, MAM, and ZA chains, respectively. Since the plasma frequency depends solely on the oxide thickness and not on the geometry of the junctions, these small variations are attributed to fluctuations in the oxidation process (see Appendix~\ref{app:Fabrication}).

We also obtain fundamental mode frequencies \( \omega_1/2\pi = 6.66 \,\mathrm{GHz} \), \( 3.79 \,\mathrm{GHz} \), and \( 1.40 \,\mathrm{GHz} \) for the Reference, MAM, and ZA chains, respectively. By injecting the total inductance \( L_{\mathrm{tot}} \) obtained from room-temperature resistance measurements (see Section~\ref{sec:dc_measurements}) into the analytical expression for \( \omega_1 \), given by Eq.~\eqref{eq:omega_1}, we estimate characteristic impedances of \( Z_{c}^{\text{Ref}} = 1860 \pm 186\,\Omega \) and \( Z_{c}^{\text{MAM}} = 4250 \pm 425\,\Omega \). Here, the uncertainty is primarily due to the systematic error in estimating \( L_J \) from resistance measurements.
Using instead the values of \( C_g \) obtained from electrostatic simulations, we find consistent impedance estimates of \( Z_{c}^{\text{Ref}} = 2340 \pm 234\,\Omega \) and \( Z_{c}^{\text{MAM}} = 4100 \pm 410\,\Omega \), , with the uncertainty primarily stemming from imperfect knowledge of the geometry used to estimate \( C_g\) in the simulations. The good agreement between the two approaches—within 20\% and 4\%, respectively—provides confidence in the robustness of our parameter extraction methods.
For the ZA chain, a room-temperature estimate of \( L_J \) is not available. However, using the impedance \( Z_c = 16{,}200 \pm 1{,}620\,\Omega \) extracted via the capacitance-based method, we deduce a Josephson inductance of \( L_J \simeq 4.75\,\mathrm{nH} \). This value is consistent with the mean junction area \( \mathcal{A} \sim\) \SI{0.7}{\micro\metre\squared}  obtained from SEM observations, and the typical value of \( L_J = (4\,\mathrm{nH} \times \) \SI{1}{\micro\metre\squared}\(/\mathcal{A}) \sim 5.7\,\mathrm{nH} \), routinely measured for our oxidation recipe.
The observed impedance scaling between the Reference and MAM chains confirms the expected \( Z_c \propto \sqrt{n} \) behavior with the number of junctions in each unit cell for similar geometries. Strikingly, the impedance of the ZA chain reaches \( Z_c \sim 16\,\mathrm{k}\Omega \), significantly exceeding the superconducting resistance quantum \( R_Q = h/(2e)^2 \simeq 6.5\,\mathrm{k}\Omega \). This places the ZA chain firmly in the high-impedance regime.

\section{Microwave mode losses}
\label{sec:losses}

In this section, we characterize the chain mode losses. We focus on the modes whose frequencies lie within the bandwidth of our High Electron Mobility Transistor (HEMT) amplifier. We measure the complex reflection coefficient $S_\mathrm{11}$($\omega$) of the chain resonances using a Vector Network Analyzer (VNA). In the case of the ZA chain, only the third mode at 4.074 GHz was observed, although more modes have frequency in the detection bandwidth. Our interpretation is that those modes have insufficient coupling to be detected. 

\begin{figure}[t]
    \includegraphics[width=0.46\textwidth]{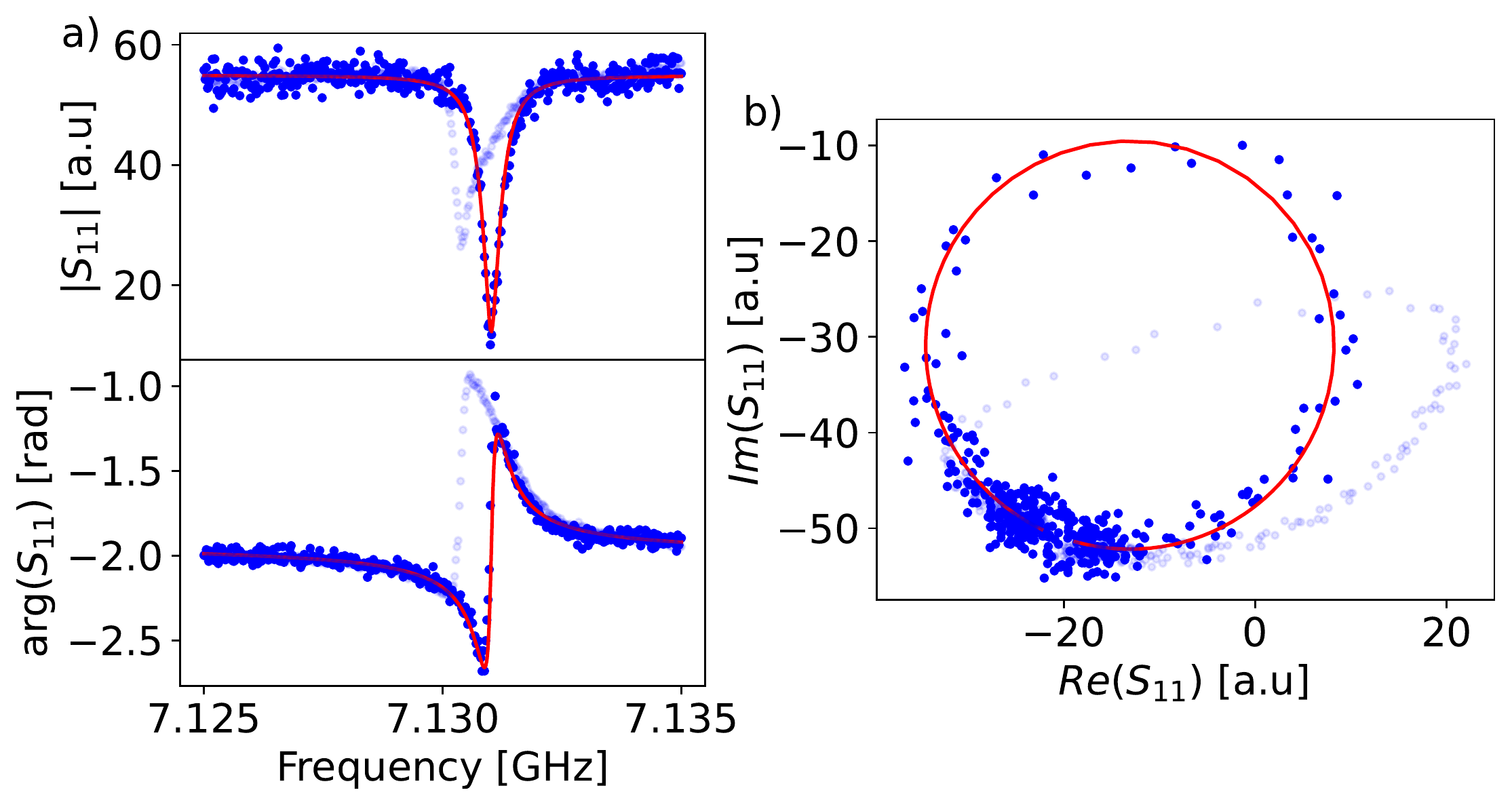}
    \caption{a) Amplitude and phase of the signal reflected by the second mode for the MAM chain with 4 junction per site. Solid blue points correspond to the linear response at low power, while transparent points indicate the distorted response at high power. Solid red line represent the fit b) Complex reflection coefficient $S_\mathrm{11}$ represented in the complex plane}
    \label{fig:figure4}
\end{figure}

The chains being composed of Josephson junctions, non linear effects are expected at high enough pump power. To ensure that we operate the junctions in the linear regime, we carefully calibrate the probing power, as shown in Fig.~\ref{fig:figure4}. The measured complex reflection coefficient of each resonance is  fitted by
\begin{equation}
    S_\mathrm{11}(\omega) = 1 - 2\frac{Re(\kappa_\mathrm{c})(1+i\tan\phi_0)}{\kappa_\mathrm{c}+ \kappa_\mathrm{i} + 2i(\omega - \omega_\mathrm{0})},
    \label{Eq:S11}
\end{equation}
where \( \kappa_\mathrm{c} \) (respectively \( \kappa_\mathrm{i} \)) is the coupling rate (respectively the internal loss rate), \( \omega_0 \) is the resonance frequency, and \( \phi_0 \) accounts for the impedance mismatch between the resonator and the input line \cite{probst2015efficient}.

From the fit, one could in principle extract both the coupling quality factor $Q_\mathrm{c} = \omega_0 / \kappa_\mathrm{c}$ and the internal quality factor $Q_\mathrm{i} = \omega_0 / \kappa_\mathrm{i}$. However, due to Fano interference effects arising from standing waves caused by signal leakage through the circulator and reflections in the measurement setup~\cite{rieger_fano_2023}, Eq.~\ref{Eq:S11} does not accurately model the measured data. As a result, it is not possible to reliably disentangle $\kappa_\mathrm{c}$ and $\kappa_\mathrm{i}$ directly from the fit. Only the total decay rate $\kappa = \kappa_\mathrm{c} + \kappa_\mathrm{i}$, and thus the total quality factor, can be extracted from the resonance linewidth. Therefore, we can only provide a lower bound for the internal quality factor.
The corresponding values are listed in Table~\ref{tab:Q_factor}.

We note that these lower bounds on the internal quality factor are inferior to values reported in the literature: Masluk et al. \cite{masluk_microwave_2012} reported a $Q_\mathrm{i} \approx$  56000 for a 160-junction chain and Kuzmin et al. \cite{kuzmin_quantum_2019} reported $Q_\mathrm{i}$ up to $\approx$ 90000 for a 33000-junction chain . Importantly, we find that the total quality factor of the reference chain, with only one junction per site, is of the same order of magnitude as that of the stacked configuration. Since the coupling geometry of the two chains is comparable, this result suggests that the use of stacked junctions does not inherently degrade the internal quality factors.

\begin{table}[!ht]
\begin{tabular}{|c c c c c|} 
\hline
Device & mode & freq. & $Q_\mathrm{tot}$ &\\
& \# & [GHz] & ~ & \\
\hline
MAM & 1& 6.266 & 6533 & \\
1 jct/site & ~ & ~ & ~ &\\
\hline
 MAM & 1 & 3.846 & 3848 &\\
 4 jct/site & 2 & 7.131 & 12 033 & \\
 & 3 & 9.668 & 7598 & \\
\hline
ZA & 3& 4.074 & 3318 & \\
9 jct/site & ~ & ~ &  ~ &\\
\hline
\end{tabular}
\caption{Properties of the modes that are directly accessible with VNA measurements}
\label{tab:Q_factor}
\end{table}

\section{Conclusion}
\label{sec:discussion}

We demonstrated two techniques for stacking Josephson junctions vertically. We showed that the characteristic impedance of a linear chain of stacks scales with the square root of the number of junctions in each stack. This observation confirms the validity of a lumped-element model in which all the junctions within a given stack share a constant stray capacitance to ground. 

Currently, chain eigenmodes do not reach state-of-the-art internal quality factors. However, we showed that this does not arise from the stacking technique. We believe that this limitation can be overcome through improved fabrication methods, such as introducing an HF cleaning step to remove niobium and silicon oxide residues before aluminum evaporation—a technique now routinely employed to fabricate highly coherent superconducting qubits \cite{bland_2d_2025,tuokkola_methods_2024}.

Compared to previous approaches used for Josephson junction stacking \cite{kohlstedt_preparation_1993,blamire_characteristics_1989}, our techniques rely on a liftoff process rather than etching. This allows stacked junction chains to be integrated into circuits after other fabrication steps, improving the flexibility of the overall fabrication process.

Both the ZA and MAM techniques have distinct advantages and drawbacks. While the MAM technique enables fabrication in a single evaporation process, it inherently limits the achievable number of stacked layers due to clogging. On the other hand, the ZA technique allows for a larger number of junctions per stack. It also offers greater flexibility to fit in arbitrary geometries and integrates well with various circuit architecture but requires two separate evaporation steps, with an intermediate exposure to air. However, a recent study \cite{wang_achieving_2024} has demonstrated that this process does not negatively affect the coherence of quantum circuits. Moreover, with a thicker resist layer, stacks of one to two micrometers in height could be reached, with few hundreds of junctions per stacks. The effect of stack inhomogeneity due to clogging of the mask by aluminum growth is not considered in this paper and will become more important the taller the junction tower. However, as the chain is intended to operate at low power where the junction response is linear, this inhomogeneity should not play a significant role.

We showed that the chain fabricated using the ZA technique—comprising $N_{\mathrm{ZA}} = 138$ stacks of $n = 9$ junctions—achieves a characteristic impedance of $Z_c \sim 16\,\mathrm{k\Omega}$ for a total inductance of $L_{\mathrm{tot}} = 5.9\,\mathrm{\mu H}$. The maximum impedance $Z(\omega)$ of the chain reached in the linear regime is bounded by $Z(\omega_1) = L_{tot}\omega_1 = \pi Z_c \sim 50$ $k \Omega$. This value is four times lower than previously reported state-of-the-art values~\cite{pechenezhskiy_superconducting_2020}.  
However, by increasing the number of junctions per stack—and ultimately reducing the chain to a single, stack—our approach has the potential to surpass existing hyperinductance implementations and unlock a new class of quantum circuits.

\subsection*{Acknowledgements}

We are grateful to Zaki Leghtas and Philippe Campagne-Ibarcq for generously granting us access to their dilution refrigerator for part of the experimental campaign. We also thank Marcelo Goffman, Hugues Pothier, Antoine Heidmann, Pierre-François Cohadon, and Tristan Briant for insightful discussions.

This work was supported by the Agence Nationale de la Recherche under projects \textsc{MecaFlux} (ANR-21-CE47-0011) \textsc{Ferbo} (ANR-23-CE47-0004), and RobustSuperQ (ANR-22-PETQ-0003), by the Région Île-de-France through the DIM \textsc{Sirteq} programme (CryoParis project), and by Sorbonne Université through the \textsc{HyQuTech} “Emergence” programme. K.\,G. acknowledges support from the Quantum Information Center Sorbonne (QICS) doctoral fellowship, and H.\,P. is funded by the CNRS–University of Arizona joint Ph.D. programme.

\bibliography{biblio.bib}

\providecommand{\noopsort}[1]{}
\begin{thebibliography}{40}%
\makeatletter
\providecommand \@ifxundefined [1]{%
 \@ifx{#1\undefined}
}%
\providecommand \@ifnum [1]{%
 \ifnum #1\expandafter \@firstoftwo
 \else \expandafter \@secondoftwo
 \fi
}%
\providecommand \@ifx [1]{%
 \ifx #1\expandafter \@firstoftwo
 \else \expandafter \@secondoftwo
 \fi
}%
\providecommand \natexlab [1]{#1}%
\providecommand \enquote  [1]{``#1''}%
\providecommand \bibnamefont  [1]{#1}%
\providecommand \bibfnamefont [1]{#1}%
\providecommand \citenamefont [1]{#1}%
\providecommand \href@noop [0]{\@secondoftwo}%
\providecommand \href [0]{\begingroup \@sanitize@url \@href}%
\providecommand \@href[1]{\@@startlink{#1}\@@href}%
\providecommand \@@href[1]{\endgroup#1\@@endlink}%
\providecommand \@sanitize@url [0]{\catcode `\\12\catcode `\$12\catcode `\&12\catcode `\#12\catcode `\^12\catcode `\_12\catcode `\%12\relax}%
\providecommand \@@startlink[1]{}%
\providecommand \@@endlink[0]{}%
\providecommand \url  [0]{\begingroup\@sanitize@url \@url }%
\providecommand \@url [1]{\endgroup\@href {#1}{\urlprefix }}%
\providecommand \urlprefix  [0]{URL }%
\providecommand \Eprint [0]{\href }%
\providecommand \doibase [0]{https://doi.org/}%
\providecommand \selectlanguage [0]{\@gobble}%
\providecommand \bibinfo  [0]{\@secondoftwo}%
\providecommand \bibfield  [0]{\@secondoftwo}%
\providecommand \translation [1]{[#1]}%
\providecommand \BibitemOpen [0]{}%
\providecommand \bibitemStop [0]{}%
\providecommand \bibitemNoStop [0]{.\EOS\space}%
\providecommand \EOS [0]{\spacefactor3000\relax}%
\providecommand \BibitemShut  [1]{\csname bibitem#1\endcsname}%
\let\auto@bib@innerbib\@empty
%</preamble>
\bibitem [{\citenamefont {Manucharyan}\ \emph {et~al.}(2009)\citenamefont {Manucharyan}, \citenamefont {Koch}, \citenamefont {Glazman},\ and\ \citenamefont {Devoret}}]{manucharyan_fluxonium_2009}%
  \BibitemOpen
  \bibfield  {author} {\bibinfo {author} {\bibfnamefont {V.~E.}\ \bibnamefont {Manucharyan}}, \bibinfo {author} {\bibfnamefont {J.}~\bibnamefont {Koch}}, \bibinfo {author} {\bibfnamefont {L.}~\bibnamefont {Glazman}},\ and\ \bibinfo {author} {\bibfnamefont {M.}~\bibnamefont {Devoret}},\ }\bibfield  {title} {\bibinfo {title} {Fluxonium: single cooper pair circuit free of charge offsets},\ }\href {https://doi.org/10.1126/science.1175552} {\bibfield  {journal} {\bibinfo  {journal} {Science}\ }\textbf {\bibinfo {volume} {326}},\ \bibinfo {pages} {113} (\bibinfo {year} {2009})},\ \Eprint {https://arxiv.org/abs/0906.0831} {0906.0831} \BibitemShut {NoStop}%
\bibitem [{\citenamefont {Pop}\ \emph {et~al.}(2014)\citenamefont {Pop}, \citenamefont {Geerlings}, \citenamefont {Catelani}, \citenamefont {Schoelkopf}, \citenamefont {Glazman},\ and\ \citenamefont {Devoret}}]{pop_coherent_2014}%
  \BibitemOpen
  \bibfield  {author} {\bibinfo {author} {\bibfnamefont {I.~M.}\ \bibnamefont {Pop}}, \bibinfo {author} {\bibfnamefont {K.}~\bibnamefont {Geerlings}}, \bibinfo {author} {\bibfnamefont {G.}~\bibnamefont {Catelani}}, \bibinfo {author} {\bibfnamefont {R.~J.}\ \bibnamefont {Schoelkopf}}, \bibinfo {author} {\bibfnamefont {L.~I.}\ \bibnamefont {Glazman}},\ and\ \bibinfo {author} {\bibfnamefont {M.~H.}\ \bibnamefont {Devoret}},\ }\bibfield  {title} {\bibinfo {title} {Coherent suppression of electromagnetic dissipation due to superconducting quasiparticles},\ }\href {https://doi.org/10.1038/nature13017} {\bibfield  {journal} {\bibinfo  {journal} {Nature}\ }\textbf {\bibinfo {volume} {508}},\ \bibinfo {pages} {369} (\bibinfo {year} {2014})},\ \bibinfo {note} {publisher: Nature Publishing Group}\BibitemShut {NoStop}%
\bibitem [{\citenamefont {Somoroff}\ \emph {et~al.}(2023)\citenamefont {Somoroff}, \citenamefont {Ficheux}, \citenamefont {Mencia}, \citenamefont {Xiong}, \citenamefont {Kuzmin},\ and\ \citenamefont {Manucharyan}}]{somoroff_millisecond_2023}%
  \BibitemOpen
  \bibfield  {author} {\bibinfo {author} {\bibfnamefont {A.}~\bibnamefont {Somoroff}}, \bibinfo {author} {\bibfnamefont {Q.}~\bibnamefont {Ficheux}}, \bibinfo {author} {\bibfnamefont {R.~A.}\ \bibnamefont {Mencia}}, \bibinfo {author} {\bibfnamefont {H.}~\bibnamefont {Xiong}}, \bibinfo {author} {\bibfnamefont {R.}~\bibnamefont {Kuzmin}},\ and\ \bibinfo {author} {\bibfnamefont {V.~E.}\ \bibnamefont {Manucharyan}},\ }\bibfield  {title} {\bibinfo {title} {Millisecond coherence in a superconducting qubit},\ }\href {https://doi.org/10.1103/PhysRevLett.130.267001} {\bibfield  {journal} {\bibinfo  {journal} {Physical Review Letters}\ }\textbf {\bibinfo {volume} {130}},\ \bibinfo {pages} {267001} (\bibinfo {year} {2023})},\ \bibinfo {note} {publisher: American Physical Society}\BibitemShut {NoStop}%
\bibitem [{\citenamefont {Koch}\ \emph {et~al.}(2009)\citenamefont {Koch}, \citenamefont {Manucharyan}, \citenamefont {Devoret},\ and\ \citenamefont {Glazman}}]{koch_charging_2009}%
  \BibitemOpen
  \bibfield  {author} {\bibinfo {author} {\bibfnamefont {J.}~\bibnamefont {Koch}}, \bibinfo {author} {\bibfnamefont {V.}~\bibnamefont {Manucharyan}}, \bibinfo {author} {\bibfnamefont {M.~H.}\ \bibnamefont {Devoret}},\ and\ \bibinfo {author} {\bibfnamefont {L.~I.}\ \bibnamefont {Glazman}},\ }\bibfield  {title} {\bibinfo {title} {Charging effects in the inductively shunted josephson junction},\ }\href {https://doi.org/10.1103/PhysRevLett.103.217004} {\bibfield  {journal} {\bibinfo  {journal} {Physical Review Letters}\ }\textbf {\bibinfo {volume} {103}},\ \bibinfo {pages} {217004} (\bibinfo {year} {2009})},\ \bibinfo {note} {publisher: American Physical Society}\BibitemShut {NoStop}%
\bibitem [{\citenamefont {Kitaev}(2006)}]{kitaev_protected_2006}%
  \BibitemOpen
  \bibfield  {author} {\bibinfo {author} {\bibfnamefont {A.}~\bibnamefont {Kitaev}},\ }\href {https://doi.org/10.48550/arXiv.cond-mat/0609441} {\bibinfo {title} {Protected qubit based on a superconducting current mirror}} (\bibinfo {year} {2006}),\ \bibinfo {note} {arXiv:cond-mat/0609441}\BibitemShut {NoStop}%
\bibitem [{\citenamefont {Brooks}\ \emph {et~al.}(2013)\citenamefont {Brooks}, \citenamefont {Kitaev},\ and\ \citenamefont {Preskill}}]{brooks_protected_2013}%
  \BibitemOpen
  \bibfield  {author} {\bibinfo {author} {\bibfnamefont {P.}~\bibnamefont {Brooks}}, \bibinfo {author} {\bibfnamefont {A.}~\bibnamefont {Kitaev}},\ and\ \bibinfo {author} {\bibfnamefont {J.}~\bibnamefont {Preskill}},\ }\bibfield  {title} {\bibinfo {title} {Protected gates for superconducting qubits},\ }\href {https://doi.org/10.1103/PhysRevA.87.052306} {\bibfield  {journal} {\bibinfo  {journal} {Physical Review A}\ }\textbf {\bibinfo {volume} {87}},\ \bibinfo {pages} {052306} (\bibinfo {year} {2013})},\ \bibinfo {note} {publisher: American Physical Society}\BibitemShut {NoStop}%
\bibitem [{\citenamefont {Groszkowski}\ \emph {et~al.}(2018)\citenamefont {Groszkowski}, \citenamefont {Paolo}, \citenamefont {Grimsmo}, \citenamefont {Blais}, \citenamefont {Schuster}, \citenamefont {Houck},\ and\ \citenamefont {Koch}}]{groszkowski_coherence_2018}%
  \BibitemOpen
  \bibfield  {author} {\bibinfo {author} {\bibfnamefont {P.}~\bibnamefont {Groszkowski}}, \bibinfo {author} {\bibfnamefont {A.~D.}\ \bibnamefont {Paolo}}, \bibinfo {author} {\bibfnamefont {A.~L.}\ \bibnamefont {Grimsmo}}, \bibinfo {author} {\bibfnamefont {A.}~\bibnamefont {Blais}}, \bibinfo {author} {\bibfnamefont {D.~I.}\ \bibnamefont {Schuster}}, \bibinfo {author} {\bibfnamefont {A.~A.}\ \bibnamefont {Houck}},\ and\ \bibinfo {author} {\bibfnamefont {J.}~\bibnamefont {Koch}},\ }\bibfield  {title} {\bibinfo {title} {Coherence properties of the 0-$\pi$ qubit},\ }\href {https://doi.org/10.1088/1367-2630/aab7cd} {\bibfield  {journal} {\bibinfo  {journal} {New Journal of Physics}\ }\textbf {\bibinfo {volume} {20}},\ \bibinfo {pages} {043053} (\bibinfo {year} {2018})},\ \bibinfo {note} {publisher: IOP Publishing}\BibitemShut {NoStop}%
\bibitem [{\citenamefont {Gyenis}\ \emph {et~al.}(2021)\citenamefont {Gyenis}, \citenamefont {Mundada}, \citenamefont {Di~Paolo}, \citenamefont {Hazard}, \citenamefont {You}, \citenamefont {Schuster}, \citenamefont {Koch}, \citenamefont {Blais},\ and\ \citenamefont {Houck}}]{gyenis_experimental_2021}%
  \BibitemOpen
  \bibfield  {author} {\bibinfo {author} {\bibfnamefont {A.}~\bibnamefont {Gyenis}}, \bibinfo {author} {\bibfnamefont {P.~S.}\ \bibnamefont {Mundada}}, \bibinfo {author} {\bibfnamefont {A.}~\bibnamefont {Di~Paolo}}, \bibinfo {author} {\bibfnamefont {T.~M.}\ \bibnamefont {Hazard}}, \bibinfo {author} {\bibfnamefont {X.}~\bibnamefont {You}}, \bibinfo {author} {\bibfnamefont {D.~I.}\ \bibnamefont {Schuster}}, \bibinfo {author} {\bibfnamefont {J.}~\bibnamefont {Koch}}, \bibinfo {author} {\bibfnamefont {A.}~\bibnamefont {Blais}},\ and\ \bibinfo {author} {\bibfnamefont {A.~A.}\ \bibnamefont {Houck}},\ }\bibfield  {title} {\bibinfo {title} {Experimental {Realization} of a {Protected} {Superconducting} {Circuit} {Derived} from the $0-\pi$ {Qubit}},\ }\href {https://doi.org/10.1103/PRXQuantum.2.010339} {\bibfield  {journal} {\bibinfo  {journal} {PRX Quantum}\ }\textbf {\bibinfo {volume} {2}},\ \bibinfo {pages} {010339} (\bibinfo {year} {2021})},\ \bibinfo {note} {publisher: American Physical Society}\BibitemShut
  {NoStop}%
\bibitem [{\citenamefont {Kalashnikov}\ \emph {et~al.}(2020)\citenamefont {Kalashnikov}, \citenamefont {Hsieh}, \citenamefont {Zhang}, \citenamefont {Lu}, \citenamefont {Kamenov}, \citenamefont {Di~Paolo}, \citenamefont {Blais}, \citenamefont {Gershenson},\ and\ \citenamefont {Bell}}]{kalashnikov_bifluxon_2020}%
  \BibitemOpen
  \bibfield  {author} {\bibinfo {author} {\bibfnamefont {K.}~\bibnamefont {Kalashnikov}}, \bibinfo {author} {\bibfnamefont {W.~T.}\ \bibnamefont {Hsieh}}, \bibinfo {author} {\bibfnamefont {W.}~\bibnamefont {Zhang}}, \bibinfo {author} {\bibfnamefont {W.-S.}\ \bibnamefont {Lu}}, \bibinfo {author} {\bibfnamefont {P.}~\bibnamefont {Kamenov}}, \bibinfo {author} {\bibfnamefont {A.}~\bibnamefont {Di~Paolo}}, \bibinfo {author} {\bibfnamefont {A.}~\bibnamefont {Blais}}, \bibinfo {author} {\bibfnamefont {M.~E.}\ \bibnamefont {Gershenson}},\ and\ \bibinfo {author} {\bibfnamefont {M.}~\bibnamefont {Bell}},\ }\bibfield  {title} {\bibinfo {title} {Bifluxon: {Fluxon}-{Parity}-{Protected} {Superconducting} {Qubit}},\ }\href {https://doi.org/10.1103/PRXQuantum.1.010307} {\bibfield  {journal} {\bibinfo  {journal} {PRX Quantum}\ }\textbf {\bibinfo {volume} {1}},\ \bibinfo {pages} {010307} (\bibinfo {year} {2020})},\ \bibinfo {note} {publisher: American Physical Society}\BibitemShut {NoStop}%
\bibitem [{\citenamefont {Bell}\ \emph {et~al.}(2016)\citenamefont {Bell}, \citenamefont {Zhang}, \citenamefont {Ioffe},\ and\ \citenamefont {Gershenson}}]{bell_spectroscopic_2016}%
  \BibitemOpen
  \bibfield  {author} {\bibinfo {author} {\bibfnamefont {M.}~\bibnamefont {Bell}}, \bibinfo {author} {\bibfnamefont {W.}~\bibnamefont {Zhang}}, \bibinfo {author} {\bibfnamefont {L.}~\bibnamefont {Ioffe}},\ and\ \bibinfo {author} {\bibfnamefont {M.}~\bibnamefont {Gershenson}},\ }\bibfield  {title} {\bibinfo {title} {Spectroscopic {Evidence} of the {Aharonov}-{Casher} {Effect} in a {Cooper} {Pair} {Box}},\ }\href {https://doi.org/10.1103/PhysRevLett.116.107002} {\bibfield  {journal} {\bibinfo  {journal} {Physical Review Letters}\ }\textbf {\bibinfo {volume} {116}},\ \bibinfo {pages} {107002} (\bibinfo {year} {2016})},\ \bibinfo {note} {publisher: American Physical Society}\BibitemShut {NoStop}%
\bibitem [{\citenamefont {Gottesman}\ \emph {et~al.}(2001)\citenamefont {Gottesman}, \citenamefont {Kitaev},\ and\ \citenamefont {Preskill}}]{gottesman_encoding_2001}%
  \BibitemOpen
  \bibfield  {author} {\bibinfo {author} {\bibfnamefont {D.}~\bibnamefont {Gottesman}}, \bibinfo {author} {\bibfnamefont {A.}~\bibnamefont {Kitaev}},\ and\ \bibinfo {author} {\bibfnamefont {J.}~\bibnamefont {Preskill}},\ }\bibfield  {title} {\bibinfo {title} {Encoding a qubit in an oscillator},\ }\href {https://doi.org/10.1103/PhysRevA.64.012310} {\bibfield  {journal} {\bibinfo  {journal} {Physical Review A}\ }\textbf {\bibinfo {volume} {64}},\ \bibinfo {pages} {012310} (\bibinfo {year} {2001})},\ \bibinfo {note} {publisher: American Physical Society}\BibitemShut {NoStop}%
\bibitem [{\citenamefont {Nathan}\ \emph {et~al.}(2024)\citenamefont {Nathan}, \citenamefont {O'Brien}, \citenamefont {Noh}, \citenamefont {Matheny}, \citenamefont {Grimsmo}, \citenamefont {Jiang},\ and\ \citenamefont {Refael}}]{nathan_self-correcting_2024}%
  \BibitemOpen
  \bibfield  {author} {\bibinfo {author} {\bibfnamefont {F.}~\bibnamefont {Nathan}}, \bibinfo {author} {\bibfnamefont {L.}~\bibnamefont {O'Brien}}, \bibinfo {author} {\bibfnamefont {K.}~\bibnamefont {Noh}}, \bibinfo {author} {\bibfnamefont {M.~H.}\ \bibnamefont {Matheny}}, \bibinfo {author} {\bibfnamefont {A.~L.}\ \bibnamefont {Grimsmo}}, \bibinfo {author} {\bibfnamefont {L.}~\bibnamefont {Jiang}},\ and\ \bibinfo {author} {\bibfnamefont {G.}~\bibnamefont {Refael}},\ }\href@noop {} {\bibinfo {title} {Self-correcting {GKP} qubit and gates in a driven–dissipative circuit}} (\bibinfo {year} {2024}),\ \Eprint {https://arxiv.org/abs/2405.05671} {arXiv:2405.05671 [quant-ph]} \BibitemShut {NoStop}%
\bibitem [{\citenamefont {Sellem}\ \emph {et~al.}(2025)\citenamefont {Sellem}, \citenamefont {Sarlette}, \citenamefont {Leghtas}, \citenamefont {Mirrahimi}, \citenamefont {Rouchon},\ and\ \citenamefont {Campagne-Ibarcq}}]{sellem_dissipative_2025}%
  \BibitemOpen
  \bibfield  {author} {\bibinfo {author} {\bibfnamefont {L.-A.}\ \bibnamefont {Sellem}}, \bibinfo {author} {\bibfnamefont {A.}~\bibnamefont {Sarlette}}, \bibinfo {author} {\bibfnamefont {Z.}~\bibnamefont {Leghtas}}, \bibinfo {author} {\bibfnamefont {M.}~\bibnamefont {Mirrahimi}}, \bibinfo {author} {\bibfnamefont {P.}~\bibnamefont {Rouchon}},\ and\ \bibinfo {author} {\bibfnamefont {P.}~\bibnamefont {Campagne-Ibarcq}},\ }\bibfield  {title} {\bibinfo {title} {Dissipative protection of a {GKP} qubit in a high-impedance superconducting circuit driven by a microwave frequency comb},\ }\href {https://doi.org/10.1103/PhysRevX.15.011011} {\bibfield  {journal} {\bibinfo  {journal} {Physical Review X}\ }\textbf {\bibinfo {volume} {15}},\ \bibinfo {pages} {011011} (\bibinfo {year} {2025})},\ \bibinfo {note} {publisher: American Physical Society}\BibitemShut {NoStop}%
\bibitem [{\citenamefont {Ranni}\ \emph {et~al.}(2023)\citenamefont {Ranni}, \citenamefont {Havir}, \citenamefont {Haldar},\ and\ \citenamefont {Maisi}}]{ranni_high_2023}%
  \BibitemOpen
  \bibfield  {author} {\bibinfo {author} {\bibfnamefont {A.}~\bibnamefont {Ranni}}, \bibinfo {author} {\bibfnamefont {H.}~\bibnamefont {Havir}}, \bibinfo {author} {\bibfnamefont {S.}~\bibnamefont {Haldar}},\ and\ \bibinfo {author} {\bibfnamefont {V.~F.}\ \bibnamefont {Maisi}},\ }\bibfield  {title} {\bibinfo {title} {High impedance {Josephson} junction resonators in the transmission line geometry},\ }\href {https://doi.org/10.1063/5.0164323} {\bibfield  {journal} {\bibinfo  {journal} {Applied Physics Letters}\ }\textbf {\bibinfo {volume} {123}},\ \bibinfo {pages} {114002} (\bibinfo {year} {2023})}\BibitemShut {NoStop}%
\bibitem [{\citenamefont {Bøttcher}\ \emph {et~al.}(2022)\citenamefont {Bøttcher}, \citenamefont {Harvey}, \citenamefont {Fallahi}, \citenamefont {Gardner}, \citenamefont {Manfra}, \citenamefont {Vool}, \citenamefont {Bartlett},\ and\ \citenamefont {Yacoby}}]{bottcher_parametric_2022}%
  \BibitemOpen
  \bibfield  {author} {\bibinfo {author} {\bibfnamefont {C.~G.~L.}\ \bibnamefont {Bøttcher}}, \bibinfo {author} {\bibfnamefont {S.~P.}\ \bibnamefont {Harvey}}, \bibinfo {author} {\bibfnamefont {S.}~\bibnamefont {Fallahi}}, \bibinfo {author} {\bibfnamefont {G.~C.}\ \bibnamefont {Gardner}}, \bibinfo {author} {\bibfnamefont {M.~J.}\ \bibnamefont {Manfra}}, \bibinfo {author} {\bibfnamefont {U.}~\bibnamefont {Vool}}, \bibinfo {author} {\bibfnamefont {S.~D.}\ \bibnamefont {Bartlett}},\ and\ \bibinfo {author} {\bibfnamefont {A.}~\bibnamefont {Yacoby}},\ }\bibfield  {title} {\bibinfo {title} {Parametric longitudinal coupling between a high-impedance superconducting resonator and a semiconductor quantum dot singlet-triplet spin qubit},\ }\href {https://doi.org/10.1038/s41467-022-32236-w} {\bibfield  {journal} {\bibinfo  {journal} {Nature Communications}\ }\textbf {\bibinfo {volume} {13}},\ \bibinfo {pages} {4773} (\bibinfo {year} {2022})},\ \bibinfo {note} {publisher: Nature Publishing Group}\BibitemShut {NoStop}%
\bibitem [{\citenamefont {Stockklauser}\ \emph {et~al.}(2017)\citenamefont {Stockklauser}, \citenamefont {Scarlino}, \citenamefont {Koski}, \citenamefont {Gasparinetti}, \citenamefont {Andersen}, \citenamefont {Reichl}, \citenamefont {Wegscheider}, \citenamefont {Ihn}, \citenamefont {Ensslin},\ and\ \citenamefont {Wallraff}}]{stockklauser_strong_2017}%
  \BibitemOpen
  \bibfield  {author} {\bibinfo {author} {\bibfnamefont {A.}~\bibnamefont {Stockklauser}}, \bibinfo {author} {\bibfnamefont {P.}~\bibnamefont {Scarlino}}, \bibinfo {author} {\bibfnamefont {J.}~\bibnamefont {Koski}}, \bibinfo {author} {\bibfnamefont {S.}~\bibnamefont {Gasparinetti}}, \bibinfo {author} {\bibfnamefont {C.~K.}\ \bibnamefont {Andersen}}, \bibinfo {author} {\bibfnamefont {C.}~\bibnamefont {Reichl}}, \bibinfo {author} {\bibfnamefont {W.}~\bibnamefont {Wegscheider}}, \bibinfo {author} {\bibfnamefont {T.}~\bibnamefont {Ihn}}, \bibinfo {author} {\bibfnamefont {K.}~\bibnamefont {Ensslin}},\ and\ \bibinfo {author} {\bibfnamefont {A.}~\bibnamefont {Wallraff}},\ }\bibfield  {title} {\bibinfo {title} {Strong {Coupling} {Cavity} {QED} with {Gate}-{Defined} {Double} {Quantum} {Dots} {Enabled} by a {High} {Impedance} {Resonator}},\ }\href {https://doi.org/10.1103/PhysRevX.7.011030} {\bibfield  {journal} {\bibinfo  {journal} {Physical Review X}\ }\textbf {\bibinfo {volume} {7}},\ \bibinfo {pages} {011030}
  (\bibinfo {year} {2017})},\ \bibinfo {note} {publisher: American Physical Society}\BibitemShut {NoStop}%
\bibitem [{\citenamefont {Landig}\ \emph {et~al.}(2018)\citenamefont {Landig}, \citenamefont {Koski}, \citenamefont {Scarlino}, \citenamefont {Mendes}, \citenamefont {Blais}, \citenamefont {Reichl}, \citenamefont {Wegscheider}, \citenamefont {Wallraff}, \citenamefont {Ensslin},\ and\ \citenamefont {Ihn}}]{landig_coherent_2018}%
  \BibitemOpen
  \bibfield  {author} {\bibinfo {author} {\bibfnamefont {A.~J.}\ \bibnamefont {Landig}}, \bibinfo {author} {\bibfnamefont {J.~V.}\ \bibnamefont {Koski}}, \bibinfo {author} {\bibfnamefont {P.}~\bibnamefont {Scarlino}}, \bibinfo {author} {\bibfnamefont {U.~C.}\ \bibnamefont {Mendes}}, \bibinfo {author} {\bibfnamefont {A.}~\bibnamefont {Blais}}, \bibinfo {author} {\bibfnamefont {C.}~\bibnamefont {Reichl}}, \bibinfo {author} {\bibfnamefont {W.}~\bibnamefont {Wegscheider}}, \bibinfo {author} {\bibfnamefont {A.}~\bibnamefont {Wallraff}}, \bibinfo {author} {\bibfnamefont {K.}~\bibnamefont {Ensslin}},\ and\ \bibinfo {author} {\bibfnamefont {T.}~\bibnamefont {Ihn}},\ }\bibfield  {title} {\bibinfo {title} {Coherent spin–photon coupling using a resonant exchange qubit},\ }\href {https://doi.org/10.1038/s41586-018-0365-y} {\bibfield  {journal} {\bibinfo  {journal} {Nature}\ }\textbf {\bibinfo {volume} {560}},\ \bibinfo {pages} {179} (\bibinfo {year} {2018})},\ \bibinfo {note} {publisher: Nature Publishing
  Group}\BibitemShut {NoStop}%
\bibitem [{\citenamefont {Scarlino}\ \emph {et~al.}(2022)\citenamefont {Scarlino}, \citenamefont {Ungerer}, \citenamefont {van Woerkom}, \citenamefont {Mancini}, \citenamefont {Stano}, \citenamefont {Müller}, \citenamefont {Landig}, \citenamefont {Koski}, \citenamefont {Reichl}, \citenamefont {Wegscheider}, \citenamefont {Ihn}, \citenamefont {Ensslin},\ and\ \citenamefont {Wallraff}}]{scarlino_situ_2022}%
  \BibitemOpen
  \bibfield  {author} {\bibinfo {author} {\bibfnamefont {P.}~\bibnamefont {Scarlino}}, \bibinfo {author} {\bibfnamefont {J.}~\bibnamefont {Ungerer}}, \bibinfo {author} {\bibfnamefont {D.}~\bibnamefont {van Woerkom}}, \bibinfo {author} {\bibfnamefont {M.}~\bibnamefont {Mancini}}, \bibinfo {author} {\bibfnamefont {P.}~\bibnamefont {Stano}}, \bibinfo {author} {\bibfnamefont {C.}~\bibnamefont {Müller}}, \bibinfo {author} {\bibfnamefont {A.}~\bibnamefont {Landig}}, \bibinfo {author} {\bibfnamefont {J.}~\bibnamefont {Koski}}, \bibinfo {author} {\bibfnamefont {C.}~\bibnamefont {Reichl}}, \bibinfo {author} {\bibfnamefont {W.}~\bibnamefont {Wegscheider}}, \bibinfo {author} {\bibfnamefont {T.}~\bibnamefont {Ihn}}, \bibinfo {author} {\bibfnamefont {K.}~\bibnamefont {Ensslin}},\ and\ \bibinfo {author} {\bibfnamefont {A.}~\bibnamefont {Wallraff}},\ }\bibfield  {title} {\bibinfo {title} {In situ {Tuning} of the {Electric}-{Dipole} {Strength} of a {Double}-{Dot} {Charge} {Qubit}: {Charge}-{Noise} {Protection} and
  {Ultrastrong} {Coupling}},\ }\href {https://doi.org/10.1103/PhysRevX.12.031004} {\bibfield  {journal} {\bibinfo  {journal} {Physical Review X}\ }\textbf {\bibinfo {volume} {12}},\ \bibinfo {pages} {031004} (\bibinfo {year} {2022})},\ \bibinfo {note} {publisher: American Physical Society}\BibitemShut {NoStop}%
\bibitem [{\citenamefont {Niepce}\ \emph {et~al.}(2019)\citenamefont {Niepce}, \citenamefont {Burnett},\ and\ \citenamefont {Bylander}}]{niepce_high_2019}%
  \BibitemOpen
  \bibfield  {author} {\bibinfo {author} {\bibfnamefont {D.}~\bibnamefont {Niepce}}, \bibinfo {author} {\bibfnamefont {J.}~\bibnamefont {Burnett}},\ and\ \bibinfo {author} {\bibfnamefont {J.}~\bibnamefont {Bylander}},\ }\bibfield  {title} {\bibinfo {title} {High kinetic inductance nbn nanowire superinductors},\ }\href {https://doi.org/10.1103/PhysRevApplied.11.044014} {\bibfield  {journal} {\bibinfo  {journal} {Physical Review Applied}\ }\textbf {\bibinfo {volume} {11}},\ \bibinfo {pages} {044014} (\bibinfo {year} {2019})},\ \bibinfo {note} {publisher: American Physical Society}\BibitemShut {NoStop}%
\bibitem [{\citenamefont {Shearrow}\ \emph {et~al.}(2018)\citenamefont {Shearrow}, \citenamefont {Koolstra}, \citenamefont {Whiteley}, \citenamefont {Earnest}, \citenamefont {Barry}, \citenamefont {Heremans}, \citenamefont {Awschalom}, \citenamefont {Shirokoff},\ and\ \citenamefont {Schuster}}]{shearrow_atomic_2018}%
  \BibitemOpen
  \bibfield  {author} {\bibinfo {author} {\bibfnamefont {A.}~\bibnamefont {Shearrow}}, \bibinfo {author} {\bibfnamefont {G.}~\bibnamefont {Koolstra}}, \bibinfo {author} {\bibfnamefont {S.~J.}\ \bibnamefont {Whiteley}}, \bibinfo {author} {\bibfnamefont {N.}~\bibnamefont {Earnest}}, \bibinfo {author} {\bibfnamefont {P.~S.}\ \bibnamefont {Barry}}, \bibinfo {author} {\bibfnamefont {F.~J.}\ \bibnamefont {Heremans}}, \bibinfo {author} {\bibfnamefont {D.~D.}\ \bibnamefont {Awschalom}}, \bibinfo {author} {\bibfnamefont {E.}~\bibnamefont {Shirokoff}},\ and\ \bibinfo {author} {\bibfnamefont {D.~I.}\ \bibnamefont {Schuster}},\ }\bibfield  {title} {\bibinfo {title} {Atomic layer deposition of titanium nitride for quantum circuits},\ }\href {https://doi.org/10.1063/1.5053461} {\bibfield  {journal} {\bibinfo  {journal} {Applied Physics Letters}\ }\textbf {\bibinfo {volume} {113}},\ \bibinfo {pages} {212601} (\bibinfo {year} {2018})}\BibitemShut {NoStop}%
\bibitem [{\citenamefont {Peltonen}\ \emph {et~al.}(2018)\citenamefont {Peltonen}, \citenamefont {Coumou}, \citenamefont {Peng}, \citenamefont {Klapwijk}, \citenamefont {Tsai},\ and\ \citenamefont {Astafiev}}]{peltonen_hybrid_2018}%
  \BibitemOpen
  \bibfield  {author} {\bibinfo {author} {\bibfnamefont {J.~T.}\ \bibnamefont {Peltonen}}, \bibinfo {author} {\bibfnamefont {P.~C. J.~J.}\ \bibnamefont {Coumou}}, \bibinfo {author} {\bibfnamefont {Z.~H.}\ \bibnamefont {Peng}}, \bibinfo {author} {\bibfnamefont {T.~M.}\ \bibnamefont {Klapwijk}}, \bibinfo {author} {\bibfnamefont {J.~S.}\ \bibnamefont {Tsai}},\ and\ \bibinfo {author} {\bibfnamefont {O.~V.}\ \bibnamefont {Astafiev}},\ }\bibfield  {title} {\bibinfo {title} {Hybrid rf {SQUID} qubit based on high kinetic inductance},\ }\href {https://doi.org/10.1038/s41598-018-27154-1} {\bibfield  {journal} {\bibinfo  {journal} {Scientific Reports}\ }\textbf {\bibinfo {volume} {8}},\ \bibinfo {pages} {10033} (\bibinfo {year} {2018})},\ \bibinfo {note} {publisher: Nature Publishing Group}\BibitemShut {NoStop}%
\bibitem [{\citenamefont {Grünhaupt}\ \emph {et~al.}(2019)\citenamefont {Grünhaupt}, \citenamefont {Spiecker}, \citenamefont {Gusenkova}, \citenamefont {Maleeva}, \citenamefont {Skacel}, \citenamefont {Takmakov}, \citenamefont {Valenti}, \citenamefont {Winkel}, \citenamefont {Rotzinger}, \citenamefont {Wernsdorfer}, \citenamefont {Ustinov},\ and\ \citenamefont {Pop}}]{grunhaupt_granular_2019}%
  \BibitemOpen
  \bibfield  {author} {\bibinfo {author} {\bibfnamefont {L.}~\bibnamefont {Grünhaupt}}, \bibinfo {author} {\bibfnamefont {M.}~\bibnamefont {Spiecker}}, \bibinfo {author} {\bibfnamefont {D.}~\bibnamefont {Gusenkova}}, \bibinfo {author} {\bibfnamefont {N.}~\bibnamefont {Maleeva}}, \bibinfo {author} {\bibfnamefont {S.~T.}\ \bibnamefont {Skacel}}, \bibinfo {author} {\bibfnamefont {I.}~\bibnamefont {Takmakov}}, \bibinfo {author} {\bibfnamefont {F.}~\bibnamefont {Valenti}}, \bibinfo {author} {\bibfnamefont {P.}~\bibnamefont {Winkel}}, \bibinfo {author} {\bibfnamefont {H.}~\bibnamefont {Rotzinger}}, \bibinfo {author} {\bibfnamefont {W.}~\bibnamefont {Wernsdorfer}}, \bibinfo {author} {\bibfnamefont {A.~V.}\ \bibnamefont {Ustinov}},\ and\ \bibinfo {author} {\bibfnamefont {I.~M.}\ \bibnamefont {Pop}},\ }\bibfield  {title} {\bibinfo {title} {Granular aluminium as a superconducting material for high-impedance quantum circuits},\ }\href {https://doi.org/10.1038/s41563-019-0350-3} {\bibfield  {journal} {\bibinfo
  {journal} {Nature Materials}\ }\textbf {\bibinfo {volume} {18}},\ \bibinfo {pages} {816} (\bibinfo {year} {2019})},\ \bibinfo {note} {publisher: Nature Publishing Group}\BibitemShut {NoStop}%
\bibitem [{\citenamefont {Gupta}\ \emph {et~al.}(2024)\citenamefont {Gupta}, \citenamefont {Winkel}, \citenamefont {Thakur}, \citenamefont {Vlaanderen}, \citenamefont {Wang}, \citenamefont {Ganjam}, \citenamefont {Frunzio},\ and\ \citenamefont {Schoelkopf}}]{gupta_low_2024}%
  \BibitemOpen
  \bibfield  {author} {\bibinfo {author} {\bibfnamefont {V.}~\bibnamefont {Gupta}}, \bibinfo {author} {\bibfnamefont {P.}~\bibnamefont {Winkel}}, \bibinfo {author} {\bibfnamefont {N.}~\bibnamefont {Thakur}}, \bibinfo {author} {\bibfnamefont {P.~v.}\ \bibnamefont {Vlaanderen}}, \bibinfo {author} {\bibfnamefont {Y.}~\bibnamefont {Wang}}, \bibinfo {author} {\bibfnamefont {S.}~\bibnamefont {Ganjam}}, \bibinfo {author} {\bibfnamefont {L.}~\bibnamefont {Frunzio}},\ and\ \bibinfo {author} {\bibfnamefont {R.~J.}\ \bibnamefont {Schoelkopf}},\ }\href {https://doi.org/10.48550/arXiv.2411.12611} {\bibinfo {title} {Low loss lumped-element inductors made from granular aluminum}} (\bibinfo {year} {2024}),\ \bibinfo {note} {arXiv:2411.12611 [quant-ph]}\BibitemShut {NoStop}%
\bibitem [{\citenamefont {Peruzzo}\ \emph {et~al.}(2020)\citenamefont {Peruzzo}, \citenamefont {Trioni}, \citenamefont {Hassani}, \citenamefont {Zemlicka},\ and\ \citenamefont {Fink}}]{M_Peruzzo_2020}%
  \BibitemOpen
  \bibfield  {author} {\bibinfo {author} {\bibfnamefont {M.}~\bibnamefont {Peruzzo}}, \bibinfo {author} {\bibfnamefont {A.}~\bibnamefont {Trioni}}, \bibinfo {author} {\bibfnamefont {F.}~\bibnamefont {Hassani}}, \bibinfo {author} {\bibfnamefont {M.}~\bibnamefont {Zemlicka}},\ and\ \bibinfo {author} {\bibfnamefont {J.~M.}\ \bibnamefont {Fink}},\ }\bibfield  {title} {\bibinfo {title} {Surpassing the resistance quantum with a geometric superinductor},\ }\href {https://doi.org/10.1103/PhysRevApplied.14.044055} {\bibfield  {journal} {\bibinfo  {journal} {Phys. Rev. Appl.}\ }\textbf {\bibinfo {volume} {14}},\ \bibinfo {pages} {044055} (\bibinfo {year} {2020})}\BibitemShut {NoStop}%
\bibitem [{\citenamefont {Peruzzo}\ \emph {et~al.}(2021)\citenamefont {Peruzzo}, \citenamefont {Hassani}, \citenamefont {Szep}, \citenamefont {Trioni}, \citenamefont {Redchenko}, \citenamefont {{\v{Z}}emli{\v{c}}ka},\ and\ \citenamefont {Fink}}]{peruzzo2021geometric}%
  \BibitemOpen
  \bibfield  {author} {\bibinfo {author} {\bibfnamefont {M.}~\bibnamefont {Peruzzo}}, \bibinfo {author} {\bibfnamefont {F.}~\bibnamefont {Hassani}}, \bibinfo {author} {\bibfnamefont {G.}~\bibnamefont {Szep}}, \bibinfo {author} {\bibfnamefont {A.}~\bibnamefont {Trioni}}, \bibinfo {author} {\bibfnamefont {E.}~\bibnamefont {Redchenko}}, \bibinfo {author} {\bibfnamefont {M.}~\bibnamefont {{\v{Z}}emli{\v{c}}ka}},\ and\ \bibinfo {author} {\bibfnamefont {J.~M.}\ \bibnamefont {Fink}},\ }\bibfield  {title} {\bibinfo {title} {Geometric superinductance qubits: Controlling phase delocalization across a single josephson junction},\ }\href@noop {} {\bibfield  {journal} {\bibinfo  {journal} {PRX Quantum}\ }\textbf {\bibinfo {volume} {2}},\ \bibinfo {pages} {040341} (\bibinfo {year} {2021})}\BibitemShut {NoStop}%
\bibitem [{\citenamefont {Randeria}\ \emph {et~al.}(2024)\citenamefont {Randeria}, \citenamefont {Hazard}, \citenamefont {Di~Paolo}, \citenamefont {Azar}, \citenamefont {Hays}, \citenamefont {Ding}, \citenamefont {An}, \citenamefont {Gingras}, \citenamefont {Niedzielski}, \citenamefont {Stickler} \emph {et~al.}}]{randeria2024dephasing}%
  \BibitemOpen
  \bibfield  {author} {\bibinfo {author} {\bibfnamefont {M.~T.}\ \bibnamefont {Randeria}}, \bibinfo {author} {\bibfnamefont {T.~M.}\ \bibnamefont {Hazard}}, \bibinfo {author} {\bibfnamefont {A.}~\bibnamefont {Di~Paolo}}, \bibinfo {author} {\bibfnamefont {K.}~\bibnamefont {Azar}}, \bibinfo {author} {\bibfnamefont {M.}~\bibnamefont {Hays}}, \bibinfo {author} {\bibfnamefont {L.}~\bibnamefont {Ding}}, \bibinfo {author} {\bibfnamefont {J.}~\bibnamefont {An}}, \bibinfo {author} {\bibfnamefont {M.}~\bibnamefont {Gingras}}, \bibinfo {author} {\bibfnamefont {B.~M.}\ \bibnamefont {Niedzielski}}, \bibinfo {author} {\bibfnamefont {H.}~\bibnamefont {Stickler}}, \emph {et~al.},\ }\bibfield  {title} {\bibinfo {title} {Dephasing in fluxonium qubits from coherent quantum phase slips},\ }\href@noop {} {\bibfield  {journal} {\bibinfo  {journal} {PRX Quantum}\ }\textbf {\bibinfo {volume} {5}},\ \bibinfo {pages} {030341} (\bibinfo {year} {2024})}\BibitemShut {NoStop}%
\bibitem [{\citenamefont {Masluk}\ \emph {et~al.}(2012)\citenamefont {Masluk}, \citenamefont {Pop}, \citenamefont {Kamal}, \citenamefont {Minev},\ and\ \citenamefont {Devoret}}]{masluk_microwave_2012}%
  \BibitemOpen
  \bibfield  {author} {\bibinfo {author} {\bibfnamefont {N.~A.}\ \bibnamefont {Masluk}}, \bibinfo {author} {\bibfnamefont {I.~M.}\ \bibnamefont {Pop}}, \bibinfo {author} {\bibfnamefont {A.}~\bibnamefont {Kamal}}, \bibinfo {author} {\bibfnamefont {Z.~K.}\ \bibnamefont {Minev}},\ and\ \bibinfo {author} {\bibfnamefont {M.~H.}\ \bibnamefont {Devoret}},\ }\bibfield  {title} {\bibinfo {title} {Microwave {Characterization} of {Josephson} {Junction} {Arrays}: {Implementing} a {Low} {Loss} {Superinductance}},\ }\href {https://doi.org/10.1103/PhysRevLett.109.137002} {\bibfield  {journal} {\bibinfo  {journal} {Physical Review Letters}\ }\textbf {\bibinfo {volume} {109}},\ \bibinfo {pages} {137002} (\bibinfo {year} {2012})},\ \bibinfo {note} {publisher: American Physical Society}\BibitemShut {NoStop}%
\bibitem [{\citenamefont {Manucharyan}(2011)}]{Manucharyan2011}%
  \BibitemOpen
  \bibfield  {author} {\bibinfo {author} {\bibfnamefont {V.~E.}\ \bibnamefont {Manucharyan}},\ }\emph {\bibinfo {title} {Superinductance}},\ \href@noop {} {\bibinfo {type} {Ph.d. thesis}},\ \bibinfo  {school} {Yale University}, \bibinfo {address} {New Haven, Connecticut} (\bibinfo {year} {2011})\BibitemShut {NoStop}%
\bibitem [{\citenamefont {Pechenezhskiy}\ \emph {et~al.}(2020)\citenamefont {Pechenezhskiy}, \citenamefont {Mencia}, \citenamefont {Nguyen}, \citenamefont {Lin},\ and\ \citenamefont {Manucharyan}}]{pechenezhskiy_superconducting_2020}%
  \BibitemOpen
  \bibfield  {author} {\bibinfo {author} {\bibfnamefont {I.~V.}\ \bibnamefont {Pechenezhskiy}}, \bibinfo {author} {\bibfnamefont {R.~A.}\ \bibnamefont {Mencia}}, \bibinfo {author} {\bibfnamefont {L.~B.}\ \bibnamefont {Nguyen}}, \bibinfo {author} {\bibfnamefont {Y.-H.}\ \bibnamefont {Lin}},\ and\ \bibinfo {author} {\bibfnamefont {V.~E.}\ \bibnamefont {Manucharyan}},\ }\bibfield  {title} {\bibinfo {title} {The superconducting quasicharge qubit},\ }\href {https://doi.org/10.1038/s41586-020-2687-9} {\bibfield  {journal} {\bibinfo  {journal} {Nature}\ }\textbf {\bibinfo {volume} {585}},\ \bibinfo {pages} {368} (\bibinfo {year} {2020})},\ \bibinfo {note} {publisher: Nature Publishing Group}\BibitemShut {NoStop}%
\bibitem [{\citenamefont {Ambegaokar}\ and\ \citenamefont {Baratoff}(1963)}]{ambegaokar1963tunneling}%
  \BibitemOpen
  \bibfield  {author} {\bibinfo {author} {\bibfnamefont {V.}~\bibnamefont {Ambegaokar}}\ and\ \bibinfo {author} {\bibfnamefont {A.}~\bibnamefont {Baratoff}},\ }\bibfield  {title} {\bibinfo {title} {Tunneling between superconductors},\ }\href@noop {} {\bibfield  {journal} {\bibinfo  {journal} {Physical review letters}\ }\textbf {\bibinfo {volume} {10}},\ \bibinfo {pages} {486} (\bibinfo {year} {1963})}\BibitemShut {NoStop}%
\bibitem [{\citenamefont {Ferguson}\ \emph {et~al.}(2007)\citenamefont {Ferguson}, \citenamefont {Clark} \emph {et~al.}}]{ferguson2007energy}%
  \BibitemOpen
  \bibfield  {author} {\bibinfo {author} {\bibfnamefont {A.}~\bibnamefont {Ferguson}}, \bibinfo {author} {\bibfnamefont {R.}~\bibnamefont {Clark}}, \emph {et~al.},\ }\bibfield  {title} {\bibinfo {title} {Energy gap measurement of nanostructured aluminium thin films for single cooper-pairdevices},\ }\href@noop {} {\bibfield  {journal} {\bibinfo  {journal} {Superconductor Science and Technology}\ }\textbf {\bibinfo {volume} {21}},\ \bibinfo {pages} {015013} (\bibinfo {year} {2007})}\BibitemShut {NoStop}%
\bibitem [{\citenamefont {Probst}\ \emph {et~al.}(2015)\citenamefont {Probst}, \citenamefont {Song}, \citenamefont {Bushev}, \citenamefont {Ustinov},\ and\ \citenamefont {Weides}}]{probst2015efficient}%
  \BibitemOpen
  \bibfield  {author} {\bibinfo {author} {\bibfnamefont {S.}~\bibnamefont {Probst}}, \bibinfo {author} {\bibfnamefont {F.}~\bibnamefont {Song}}, \bibinfo {author} {\bibfnamefont {P.~A.}\ \bibnamefont {Bushev}}, \bibinfo {author} {\bibfnamefont {A.~V.}\ \bibnamefont {Ustinov}},\ and\ \bibinfo {author} {\bibfnamefont {M.}~\bibnamefont {Weides}},\ }\bibfield  {title} {\bibinfo {title} {Efficient and robust analysis of complex scattering data under noise in microwave resonators},\ }\href@noop {} {\bibfield  {journal} {\bibinfo  {journal} {Review of Scientific Instruments}\ }\textbf {\bibinfo {volume} {86}} (\bibinfo {year} {2015})}\BibitemShut {NoStop}%
\bibitem [{\citenamefont {Rieger}\ \emph {et~al.}(2023)\citenamefont {Rieger}, \citenamefont {Günzler}, \citenamefont {Spiecker}, \citenamefont {Nambisan}, \citenamefont {Wernsdorfer},\ and\ \citenamefont {Pop}}]{rieger_fano_2023}%
  \BibitemOpen
  \bibfield  {author} {\bibinfo {author} {\bibfnamefont {D.}~\bibnamefont {Rieger}}, \bibinfo {author} {\bibfnamefont {S.}~\bibnamefont {Günzler}}, \bibinfo {author} {\bibfnamefont {M.}~\bibnamefont {Spiecker}}, \bibinfo {author} {\bibfnamefont {A.}~\bibnamefont {Nambisan}}, \bibinfo {author} {\bibfnamefont {W.}~\bibnamefont {Wernsdorfer}},\ and\ \bibinfo {author} {\bibfnamefont {I.}~\bibnamefont {Pop}},\ }\bibfield  {title} {\bibinfo {title} {Fano {Interference} in {Microwave} {Resonator} {Measurements}},\ }\href {https://doi.org/10.1103/PhysRevApplied.20.014059} {\bibfield  {journal} {\bibinfo  {journal} {Physical Review Applied}\ }\textbf {\bibinfo {volume} {20}},\ \bibinfo {pages} {014059} (\bibinfo {year} {2023})},\ \bibinfo {note} {publisher: American Physical Society}\BibitemShut {NoStop}%
\bibitem [{\citenamefont {Kuzmin}\ \emph {et~al.}(2019)\citenamefont {Kuzmin}, \citenamefont {Mencia}, \citenamefont {Grabon}, \citenamefont {Mehta}, \citenamefont {Lin},\ and\ \citenamefont {Manucharyan}}]{kuzmin_quantum_2019}%
  \BibitemOpen
  \bibfield  {author} {\bibinfo {author} {\bibfnamefont {R.}~\bibnamefont {Kuzmin}}, \bibinfo {author} {\bibfnamefont {R.}~\bibnamefont {Mencia}}, \bibinfo {author} {\bibfnamefont {N.}~\bibnamefont {Grabon}}, \bibinfo {author} {\bibfnamefont {N.}~\bibnamefont {Mehta}}, \bibinfo {author} {\bibfnamefont {Y.-H.}\ \bibnamefont {Lin}},\ and\ \bibinfo {author} {\bibfnamefont {V.~E.}\ \bibnamefont {Manucharyan}},\ }\bibfield  {title} {\bibinfo {title} {Quantum electrodynamics of a superconductor–insulator phase transition},\ }\href {https://doi.org/10.1038/s41567-019-0553-1} {\bibfield  {journal} {\bibinfo  {journal} {Nature Physics}\ }\textbf {\bibinfo {volume} {15}},\ \bibinfo {pages} {930} (\bibinfo {year} {2019})},\ \bibinfo {note} {publisher: Nature Publishing Group}\BibitemShut {NoStop}%
\bibitem [{\citenamefont {Bland}\ \emph {et~al.}(2025)\citenamefont {Bland}, \citenamefont {Bahrami}, \citenamefont {Martinez}, \citenamefont {Prestegaard}, \citenamefont {Smitham}, \citenamefont {Joshi}, \citenamefont {Hedrick}, \citenamefont {Pakpour-Tabrizi}, \citenamefont {Kumar}, \citenamefont {Jindal}, \citenamefont {Chang}, \citenamefont {Yang}, \citenamefont {Cheng}, \citenamefont {Yao}, \citenamefont {Cava}, \citenamefont {Leon},\ and\ \citenamefont {Houck}}]{bland_2d_2025}%
  \BibitemOpen
  \bibfield  {author} {\bibinfo {author} {\bibfnamefont {M.~P.}\ \bibnamefont {Bland}}, \bibinfo {author} {\bibfnamefont {F.}~\bibnamefont {Bahrami}}, \bibinfo {author} {\bibfnamefont {J.~G.~C.}\ \bibnamefont {Martinez}}, \bibinfo {author} {\bibfnamefont {P.~H.}\ \bibnamefont {Prestegaard}}, \bibinfo {author} {\bibfnamefont {B.~M.}\ \bibnamefont {Smitham}}, \bibinfo {author} {\bibfnamefont {A.}~\bibnamefont {Joshi}}, \bibinfo {author} {\bibfnamefont {E.}~\bibnamefont {Hedrick}}, \bibinfo {author} {\bibfnamefont {A.}~\bibnamefont {Pakpour-Tabrizi}}, \bibinfo {author} {\bibfnamefont {S.}~\bibnamefont {Kumar}}, \bibinfo {author} {\bibfnamefont {A.}~\bibnamefont {Jindal}}, \bibinfo {author} {\bibfnamefont {R.~D.}\ \bibnamefont {Chang}}, \bibinfo {author} {\bibfnamefont {A.}~\bibnamefont {Yang}}, \bibinfo {author} {\bibfnamefont {G.}~\bibnamefont {Cheng}}, \bibinfo {author} {\bibfnamefont {N.}~\bibnamefont {Yao}}, \bibinfo {author} {\bibfnamefont {R.~J.}\ \bibnamefont {Cava}}, \bibinfo {author} {\bibfnamefont
  {N.~P.~d.}\ \bibnamefont {Leon}},\ and\ \bibinfo {author} {\bibfnamefont {A.~A.}\ \bibnamefont {Houck}},\ }\href {https://doi.org/10.48550/arXiv.2503.14798} {\bibinfo {title} {{2D} transmons with lifetimes and coherence times exceeding 1 millisecond}} (\bibinfo {year} {2025}),\ \bibinfo {note} {arXiv:2503.14798 [quant-ph]}\BibitemShut {NoStop}%
\bibitem [{\citenamefont {Tuokkola}\ \emph {et~al.}(2024)\citenamefont {Tuokkola}, \citenamefont {Sunada}, \citenamefont {Kivijärvi}, \citenamefont {Grönberg}, \citenamefont {Kaikkonen}, \citenamefont {Vesterinen}, \citenamefont {Govenius},\ and\ \citenamefont {Möttönen}}]{tuokkola_methods_2024}%
  \BibitemOpen
  \bibfield  {author} {\bibinfo {author} {\bibfnamefont {M.}~\bibnamefont {Tuokkola}}, \bibinfo {author} {\bibfnamefont {Y.}~\bibnamefont {Sunada}}, \bibinfo {author} {\bibfnamefont {H.}~\bibnamefont {Kivijärvi}}, \bibinfo {author} {\bibfnamefont {L.}~\bibnamefont {Grönberg}}, \bibinfo {author} {\bibfnamefont {J.-P.}\ \bibnamefont {Kaikkonen}}, \bibinfo {author} {\bibfnamefont {V.}~\bibnamefont {Vesterinen}}, \bibinfo {author} {\bibfnamefont {J.}~\bibnamefont {Govenius}},\ and\ \bibinfo {author} {\bibfnamefont {M.}~\bibnamefont {Möttönen}},\ }\href {https://doi.org/10.48550/arXiv.2407.18778} {\bibinfo {title} {Methods to achieve near-millisecond energy relaxation and dephasing times for a superconducting transmon qubit}} (\bibinfo {year} {2024}),\ \bibinfo {note} {arXiv:2407.18778 [quant-ph]}\BibitemShut {NoStop}%
\bibitem [{\citenamefont {Kohlstedt}\ \emph {et~al.}(1993)\citenamefont {Kohlstedt}, \citenamefont {Hallmanns}, \citenamefont {Nevirkovets}, \citenamefont {Guggi},\ and\ \citenamefont {Heiden}}]{kohlstedt_preparation_1993}%
  \BibitemOpen
  \bibfield  {author} {\bibinfo {author} {\bibfnamefont {H.}~\bibnamefont {Kohlstedt}}, \bibinfo {author} {\bibfnamefont {G.}~\bibnamefont {Hallmanns}}, \bibinfo {author} {\bibfnamefont {I.}~\bibnamefont {Nevirkovets}}, \bibinfo {author} {\bibfnamefont {D.}~\bibnamefont {Guggi}},\ and\ \bibinfo {author} {\bibfnamefont {C.}~\bibnamefont {Heiden}},\ }\bibfield  {title} {\bibinfo {title} {Preparation and properties of {Nb}/{Al}-{AlO}/sub x//{Nb} multilayers},\ }\href {https://doi.org/10.1109/77.233939} {\bibfield  {journal} {\bibinfo  {journal} {IEEE Transactions on Applied Superconductivity}\ }\textbf {\bibinfo {volume} {3}},\ \bibinfo {pages} {2197} (\bibinfo {year} {1993})},\ \bibinfo {note} {conference Name: IEEE Transactions on Applied Superconductivity}\BibitemShut {NoStop}%
\bibitem [{\citenamefont {Blamire}\ \emph {et~al.}(1989)\citenamefont {Blamire}, \citenamefont {Somekh}, \citenamefont {Morris},\ and\ \citenamefont {Evetts}}]{blamire_characteristics_1989}%
  \BibitemOpen
  \bibfield  {author} {\bibinfo {author} {\bibfnamefont {M.}~\bibnamefont {Blamire}}, \bibinfo {author} {\bibfnamefont {R.}~\bibnamefont {Somekh}}, \bibinfo {author} {\bibfnamefont {G.}~\bibnamefont {Morris}},\ and\ \bibinfo {author} {\bibfnamefont {J.}~\bibnamefont {Evetts}},\ }\bibfield  {title} {\bibinfo {title} {Characteristics of vertically-stacked planar tunnel junction structures},\ }\href {https://doi.org/10.1109/20.92489} {\bibfield  {journal} {\bibinfo  {journal} {IEEE Transactions on Magnetics}\ }\textbf {\bibinfo {volume} {25}},\ \bibinfo {pages} {1135} (\bibinfo {year} {1989})},\ \bibinfo {note} {conference Name: IEEE Transactions on Magnetics}\BibitemShut {NoStop}%
\bibitem [{\citenamefont {Wang}\ \emph {et~al.}(2024)\citenamefont {Wang}, \citenamefont {Lu}, \citenamefont {Zhan}, \citenamefont {Ma}, \citenamefont {Wu}, \citenamefont {Sun}, \citenamefont {Deng}, \citenamefont {Bai}, \citenamefont {Bao}, \citenamefont {Chang}, \citenamefont {Gao}, \citenamefont {Gao}, \citenamefont {Gong}, \citenamefont {Hu}, \citenamefont {Hu}, \citenamefont {Ji}, \citenamefont {Ma}, \citenamefont {Mao}, \citenamefont {Song}, \citenamefont {Tang}, \citenamefont {Wang}, \citenamefont {Wang}, \citenamefont {Wang}, \citenamefont {Xia}, \citenamefont {Xu}, \citenamefont {Zhan}, \citenamefont {Zhang}, \citenamefont {Zhou}, \citenamefont {Zhu}, \citenamefont {Zhu}, \citenamefont {Zhu}, \citenamefont {Zhu}, \citenamefont {Shi}, \citenamefont {Zhao},\ and\ \citenamefont {Deng}}]{wang_achieving_2024}%
  \BibitemOpen
  \bibfield  {author} {\bibinfo {author} {\bibfnamefont {F.}~\bibnamefont {Wang}}, \bibinfo {author} {\bibfnamefont {K.}~\bibnamefont {Lu}}, \bibinfo {author} {\bibfnamefont {H.}~\bibnamefont {Zhan}}, \bibinfo {author} {\bibfnamefont {L.}~\bibnamefont {Ma}}, \bibinfo {author} {\bibfnamefont {F.}~\bibnamefont {Wu}}, \bibinfo {author} {\bibfnamefont {H.}~\bibnamefont {Sun}}, \bibinfo {author} {\bibfnamefont {H.}~\bibnamefont {Deng}}, \bibinfo {author} {\bibfnamefont {Y.}~\bibnamefont {Bai}}, \bibinfo {author} {\bibfnamefont {F.}~\bibnamefont {Bao}}, \bibinfo {author} {\bibfnamefont {X.}~\bibnamefont {Chang}}, \bibinfo {author} {\bibfnamefont {R.}~\bibnamefont {Gao}}, \bibinfo {author} {\bibfnamefont {X.}~\bibnamefont {Gao}}, \bibinfo {author} {\bibfnamefont {G.}~\bibnamefont {Gong}}, \bibinfo {author} {\bibfnamefont {L.}~\bibnamefont {Hu}}, \bibinfo {author} {\bibfnamefont {R.}~\bibnamefont {Hu}}, \bibinfo {author} {\bibfnamefont {H.}~\bibnamefont {Ji}}, \bibinfo {author} {\bibfnamefont {X.}~\bibnamefont {Ma}},
  \bibinfo {author} {\bibfnamefont {L.}~\bibnamefont {Mao}}, \bibinfo {author} {\bibfnamefont {Z.}~\bibnamefont {Song}}, \bibinfo {author} {\bibfnamefont {C.}~\bibnamefont {Tang}}, \bibinfo {author} {\bibfnamefont {H.}~\bibnamefont {Wang}}, \bibinfo {author} {\bibfnamefont {T.}~\bibnamefont {Wang}}, \bibinfo {author} {\bibfnamefont {Z.}~\bibnamefont {Wang}}, \bibinfo {author} {\bibfnamefont {T.}~\bibnamefont {Xia}}, \bibinfo {author} {\bibfnamefont {H.}~\bibnamefont {Xu}}, \bibinfo {author} {\bibfnamefont {Z.}~\bibnamefont {Zhan}}, \bibinfo {author} {\bibfnamefont {G.}~\bibnamefont {Zhang}}, \bibinfo {author} {\bibfnamefont {T.}~\bibnamefont {Zhou}}, \bibinfo {author} {\bibfnamefont {M.}~\bibnamefont {Zhu}}, \bibinfo {author} {\bibfnamefont {Q.}~\bibnamefont {Zhu}}, \bibinfo {author} {\bibfnamefont {S.}~\bibnamefont {Zhu}}, \bibinfo {author} {\bibfnamefont {X.}~\bibnamefont {Zhu}}, \bibinfo {author} {\bibfnamefont {Y.}~\bibnamefont {Shi}}, \bibinfo {author} {\bibfnamefont {H.-H.}\ \bibnamefont {Zhao}},\ and\
  \bibinfo {author} {\bibfnamefont {C.}~\bibnamefont {Deng}},\ }\href {https://doi.org/10.48550/arXiv.2405.05481} {\bibinfo {title} {Achieving millisecond coherence fluxonium through overlap {Josephson} junctions}} (\bibinfo {year} {2024}),\ \bibinfo {note} {arXiv:2405.05481 [quant-ph]}\BibitemShut {NoStop}%
\bibitem [{\citenamefont {Potts}\ \emph {et~al.}(2001)\citenamefont {Potts}, \citenamefont {Parker}, \citenamefont {Baumberg},\ and\ \citenamefont {De~Groot}}]{potts2001cmos}%
  \BibitemOpen
  \bibfield  {author} {\bibinfo {author} {\bibfnamefont {A.}~\bibnamefont {Potts}}, \bibinfo {author} {\bibfnamefont {G.}~\bibnamefont {Parker}}, \bibinfo {author} {\bibfnamefont {J.}~\bibnamefont {Baumberg}},\ and\ \bibinfo {author} {\bibfnamefont {P.}~\bibnamefont {De~Groot}},\ }\bibfield  {title} {\bibinfo {title} {Cmos compatible fabrication methods for submicron josephson junction qubits},\ }\href@noop {} {\bibfield  {journal} {\bibinfo  {journal} {IEE Proceedings-Science, Measurement and Technology}\ }\textbf {\bibinfo {volume} {148}},\ \bibinfo {pages} {225} (\bibinfo {year} {2001})}\BibitemShut {NoStop}%
\end{thebibliography}%

\newpage

\appendix

\section{Chain mode spectrum derivation}
\label{app:theory}

\begin{figure}
\includegraphics[scale=0.52]{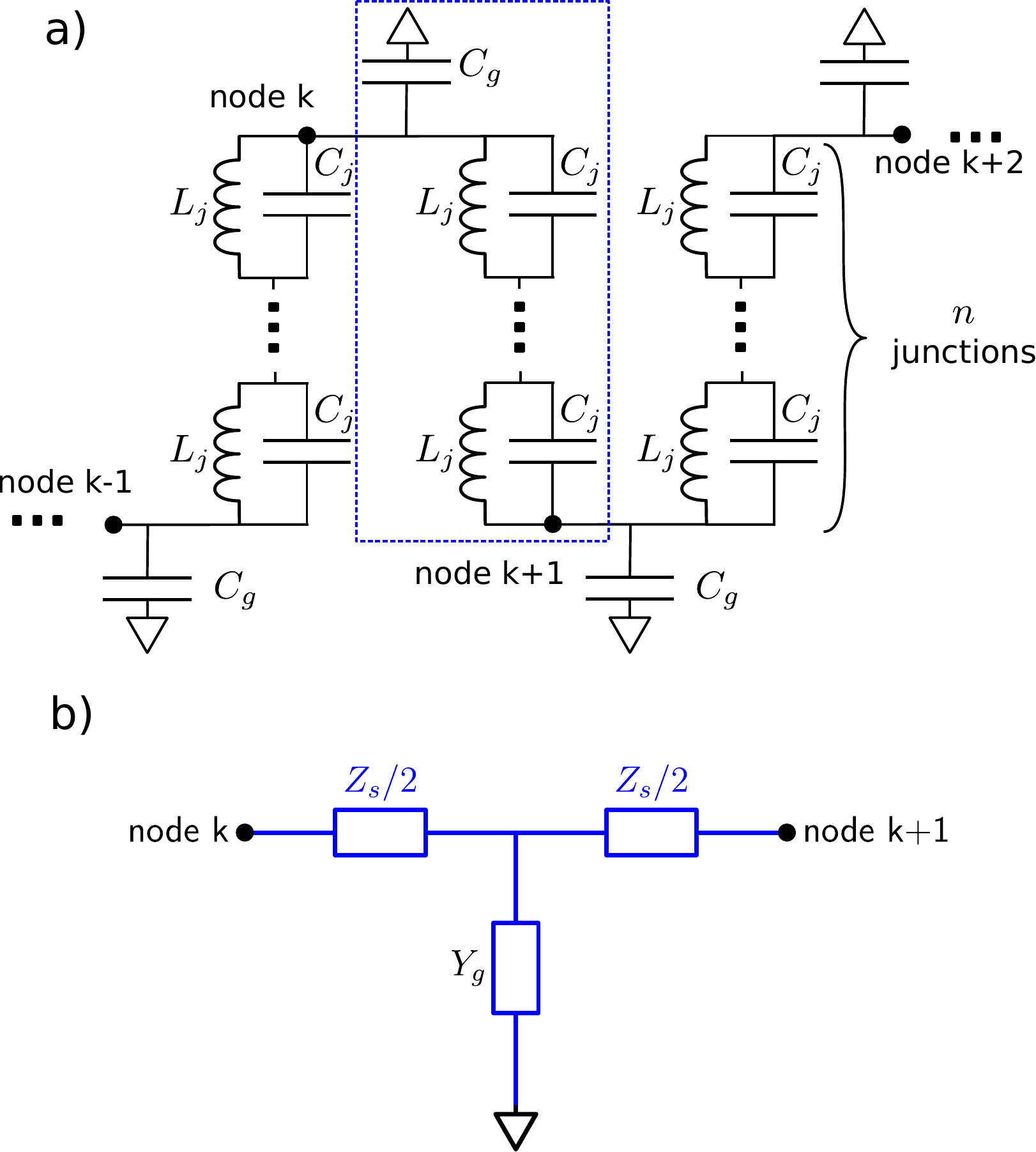}
\caption{a) Lump-element model of a chain of junction stacks. The blue dashed line represents the unit cell. b) Symmetrized unit cell of the linear junction chain. $Y_g$ is the admittance of the capacitance to ground, and $Z_s$ the impedance of one stack of $n$ junctions.}
\label{fig:unit_cell}
\end{figure}

In this appendix we derive the junction chain mode spectrum by using standard transmission line theory. 
We divide the chain into $N$ identical unit cells represented by a two-port network in the symmetrized ``T'' configuration (see Fig. \ref{fig:unit_cell}).
We start by evaluating the ABCD matrix $M_u$ of a single unit cell. We recall that the stack impedance, and admittance to ground write
\begin{equation}
Z_s = n\frac{jL_J\omega}{1-\frac{\omega^2}{\omega_p^2}}, ~~ Y_g = jC_g\omega,
\end{equation}
with $\omega_p = 1/\sqrt{L_J C_J}$. The ABCD matrix of the unit cell then writes
\begin{equation}
    M_u = \begin{pmatrix}A & B\\ C & D\end{pmatrix} = \begin{pmatrix}1 & Z_s/2\\ 0 & 1\end{pmatrix}\begin{pmatrix}1 & 0\\ Y_g & 1\end{pmatrix}\begin{pmatrix}1 & Z_s/2\\ 0 & 1\end{pmatrix}.
\end{equation}
We identify
\begin{equation}
    A=1+Z_sY_g/2, ~~ B=Z_s(1+Z_sY_g/4), ~~ C=Y_g.
\end{equation}
On can check that $A=D$, and $\det(M_u)=A^2-CB=1$. This is the matrix of a symmetric, reciprocal, and lossless transmission line unit cell. The impedance of the unit cell writes
\begin{equation}
    Z_u(\omega)\!=\!\sqrt{\frac{B}{C}}\!=\!\sqrt{\frac{nL_J}{C_g}}\frac{1}{\sqrt{1-\frac{\omega^2}{\omega_p^2}}}\sqrt{1-\frac{\omega^2/\omega_g^2}{4(1-\frac{\omega^2}{\omega_p^2})}},
    \label{Eq:impedanceABCD}
\end{equation}
with $\omega_g = 1/\sqrt{n L_J C_g}$. This expression reduces to $Z_c=\sqrt{nL_J/C_g}$ for $\omega\ll\omega_p,~ \omega_g$.

In the following, we express $M_u$ in a more convenient way. Since $|A|<1$, we can introduce the phase $\theta$ defined by $A= \cos\theta$. Since $\det M_u=1$, we have $CB=-\sin^2\theta$. From Eq.~\ref{Eq:impedanceABCD}, we get $B=jZ_u\sin\theta$ and $C=j\sin\theta/Z_u.$

The ABCD matrix of the full chain is \begin{align}
M=(M_u)^N &= \begin{pmatrix}\cos\theta & jZ_u\sin\theta\\j\sin\theta/Z_u & \cos\theta \end{pmatrix}^N\\
&= \begin{pmatrix}\cos\!{(N\theta)} & jZ_u\sin\!{(N\theta)}\\j\sin\!{(N\theta)}/Z_u & \cos\!{(N\theta)} \end{pmatrix},
\label{Eq:M}
\end{align}
where the last step can be obtained by exponentiation of the diagonal form of $M_u$, whose eigenvalues are $e^{i\theta}$ and $e^{-i\theta}$.

We now impose boundary conditions on the chain, assuming it is open at both ends, as described in the reference article. In the ABCD matrix formalism, this implies that both the input and output currents are zero, i.e.,

\begin{align}
    \begin{pmatrix} V_{in} \\ I_{in} = 0 \end{pmatrix} = M \begin{pmatrix} V_{out} \\ I_{out} = 0 \end{pmatrix}.
\end{align}
This leads to the condition 
\begin{align}
\forall V_{out}, \quad j\frac{\sin\!{(N\theta)}}{Z_u} &= 0, ~\text{such that}\\
\sin\!{(N\theta)} &= 0.
\end{align}
The Bloch phase of this periodic chain satisfy 
\begin{equation}
\theta_m = \frac{m\pi}{N}, \quad \text{with} \quad m \in \{1, \dots, N-1\}.
\end{equation}
Substituting this condition in the unit cell dispersion relation
\begin{equation}
A=\cos\theta(\omega) = 1 - \frac{\omega^2/\omega_g^2}{2(1-\omega^2/\omega_p^2)},
\end{equation}
we get the eigenmodes angular frequencies
\begin{equation}
    \omega_m = \omega_p \sqrt{\frac{2(1 - \cos\theta_m)}{(\frac{\omega_p}{\omega_g})^2 + 2(1 - \cos\theta_m)}}.
    \label{Eq:eigenmodes}
\end{equation}

 At low mode index, $N\gg m$, leading to $1-\cos\theta_m\simeq \theta_m^2/2$. Additionally, if $\omega_p/ \omega_g\gg \pi/N$, that is to say if $\sqrt{C_g/C_J}\gg \pi/N\!\sqrt{n}$
\begin{equation}
\omega_m\simeq \frac{m\pi}{N}\omega_g.
\end{equation}
In that case, the low frequency derivative of Eq.~\ref{Eq:eigenmodes} provides a direct measure of $\omega_g$.

At high mode index, there always exists a small integer $p$ such that $m=N-p$, and $\theta_m=\pi-p\pi/N$, with $p\pi/N\ll 1$. In that case, we have $\cos\theta_m\simeq -(1-(p\pi/N)^2/2)$, and the mode frequencies asymptote at large $m$ is 
\begin{equation}
     \omega_m\rightarrow
 \frac{2\omega_g}{\sqrt{1 + 4\tfrac{\omega_g^2}{\omega_p^2}}}.
\end{equation}
Finally, in the limit $\omega_g\gg\omega_p$, we get
\begin{equation}
     \omega_m\rightarrow\omega_p.
 \end{equation} The large index asymptote provides the plasma frequency.

\section{Fabrication}
\label{app:Fabrication}

Our fabrication methods relies on a lift-off process. For both techniques, we create a bilayer of MMA-PMMA and we perform a series of aluminum depositions, followed by a 10-minute oxidation at an oxygen pressure of 200 mbar. We now turn to the specific characteristics of each fabrication method.

\subsection{Multi Angle Manhattan process}

This technique is derived from the well known Manhattan technique \cite{potts2001cmos} where the great height of the PMMA layer on top of a MMA layer, as well as perpendicular \textit{streets} in the resists, are used to create Josephson junctions. The orthogonal pattern, done by electron-beam lithography, allows one to deposit aluminum in one direction only, the other one being in the shadows of the PMMA walls when choosing the right zenithal angle. The classic Manhattan technique can be finished by oxidizing those first leads and then rotating the chip by $90^\circ$, now giving access to the second direction only. 

The MAM technique uses the same basic ingredients, high PMMA walls above a MMA layer and perpendicular streets but adds a twist to it : the non symmetric shape of the Junctions (here triangles) gives access to multiple new interesting angles and enables junction stacking without ever needing to get the sample out of the evaporator. We have a pair of planar and zenithal angles that define the first horizontal leads, and then other values for the second, third and fourth aluminum layers with the fifth one, connecting the whole structure and finishing the circuit, being vertical. These angles give specific deposition areas so that each layer only covers the previous one and thus force electrons to go through each and every junction in the stack successively. They were found thanks to a home-built Python script which emulates the evaporation process. Here is the set of angles we used in our fabrication, we give ($\theta$, $\phi$) with $\theta$ being the planar angle in the XY-plane of the chip and $\phi$ the zenithal angle calculated from the Z-direction: I) ($180^\circ$, $49^\circ$) ; II) ($150^\circ$, $48^\circ$) ; III) ($127.5^\circ$, $50^\circ$) ; IV) ($113.5^\circ$, $53.5^\circ$) ; V) ($90^\circ$, $55^\circ$). The chip then finally goes through a regular lift-off in an acetone hot bath.

With this Manhattan multi-angle evaporation method, the differents angles lead to variations in overlap between layers and for the 4-junction stacks discussed in this paper, the junction area inside a stack typically ranges from \SI{1.65}{\micro\metre\squared} to \SI{2.3}{\micro\metre\squared}.

\subsection{Zero Angle process}

The zero-angle technique is based on two evaporation steps separated by an exposure to air. Prior to junction fabrication, we deposit aluminum pads spaced at regular intervals. These pads serve as current leads for the bases of the junction towers.

The chip is then coated with a bilayer resist (MMA/PMMA), and square openings are patterned at both ends of each aluminum pad using electron-beam lithography. The chip is subsequently loaded into an electron-beam evaporator, where the native aluminum oxide is removed by an ion milling step. The junction towers are then deposited by aluminum evaporation at normal incidence. Following deposition, the aluminum on top of the resist is removed through a standard lift-off process using an acetone bath. Due to progressive mask clogging caused by aluminum deposition on the resist sidewalls, the resulting junction towers exhibit a pyramidal profile (see Figure~\ref{fig:blender}c). The typical top angle of the pyramid is measured to be $\theta = 12.8^\circ$. As a result, the difference in area between the square base of the pyramid (with side length $l$) and a horizontal cross-section at height $d$ is given by $\Delta a = 4ld\tan\theta - 4d^2\tan^2\theta$. In our chain, the resist aperture has an initial area of \SI{1}{\micro\metre\squared}, and the multilayer stack consists of 9 layers, each spaced by 20 nm. Accordingly, the resulting area reduction between the top and the base layer is $\Delta a = $ \SI{0.18}{\micro\metre\squared}.

To electrically reconnect the junction stacks and properly define the chain, recontact is implemented using air bridges. The process begins by coating the chip with a PMMA layer, in which square openings are patterned at the center of each junction stack to define the bridge piers. The resist is then reflowed by heating the chip on a hot plate. A PMGI sacrificial layer is subsequently spinned onto the chip to protect the reflowed resist, followed by a new PMMA layer. Rectangular windows are then defined by electron-beam lithography to form the bridge decks, and the PMGI layer is removed using a basic solution that does not develop the PMMA.A. The chip is then reloaded into the evaporation chamber. Prior to metal deposition, an ion milling step is performed to remove the native $\mathrm{Al_20_3}$ layer formed during the exposure to air. Subsequently, 800 nm of aluminum is deposited to form the air bridges, and a final lift-off process completes the fabrication.

\end{document}